\documentclass[twocolumn,letterpaper,aps,prc,longbibliography,superscriptaddress,showpacs,nofootinbib,floatfix]{revtex4-1}

\usepackage{graphicx}	
\usepackage{amsmath,amssymb,amsfonts}
\usepackage{xspace}	

\newcommand{\pt}{\mbox{$p_T$}\xspace}

\newcommand{\sqsn}{\mbox{$\sqrt{s_{_{NN}}}$}\xspace}

\def\Cher{\mbox{$\check{\rm C}$erenkov }}

\newcommand{\mean}[1]{\left\langle #1 \right\rangle}

\newcommand{\eps}{\mbox{${\varepsilon}$}\xspace}

\begin{document}

\title{
Measurements of directed, elliptic, and triangular flow in Cu$+$Au 
collisions at $\sqrt{s_{_{NN}}}=200$~GeV}

\newcommand{\abilene}{Abilene Christian University, Abilene, Texas 79699, USA}
\newcommand{\augie}{Department of Physics, Augustana University, Sioux Falls, South Dakota 57197, USA}
\newcommand{\banaras}{Department of Physics, Banaras Hindu University, Varanasi 221005, India}
\newcommand{\barc}{Bhabha Atomic Research Centre, Bombay 400 085, India}
\newcommand{\baruch}{Baruch College, City University of New York, New York, New York, 10010 USA}
\newcommand{\bnlcoll}{Collider-Accelerator Department, Brookhaven National Laboratory, Upton, New York 11973-5000, USA}
\newcommand{\bnlphys}{Physics Department, Brookhaven National Laboratory, Upton, New York 11973-5000, USA}
\newcommand{\caucr}{University of California-Riverside, Riverside, California 92521, USA}
\newcommand{\charlesczech}{Charles University, Ovocn\'{y} trh 5, Praha 1, 116 36, Prague, Czech Republic}
\newcommand{\chonbuk}{Chonbuk National University, Jeonju, 561-756, Korea}
\newcommand{\ciae}{Science and Technology on Nuclear Data Laboratory, China Institute of Atomic Energy, Beijing 102413, P.~R.~China}
\newcommand{\cns}{Center for Nuclear Study, Graduate School of Science, University of Tokyo, 7-3-1 Hongo, Bunkyo, Tokyo 113-0033, Japan}
\newcommand{\colorado}{University of Colorado, Boulder, Colorado 80309, USA}
\newcommand{\columbia}{Columbia University, New York, New York 10027 and Nevis Laboratories, Irvington, New York 10533, USA}
\newcommand{\czechtech}{Czech Technical University, Zikova 4, 166 36 Prague 6, Czech Republic}
\newcommand{\elte}{ELTE, E{\"o}tv{\"o}s Lor{\'a}nd University, H-1117 Budapest, P{\'a}zm{\'a}ny P.~s.~1/A, Hungary}
\newcommand{\ewha}{Ewha Womans University, Seoul 120-750, Korea}
\newcommand{\fsu}{Florida State University, Tallahassee, Florida 32306, USA}
\newcommand{\gsu}{Georgia State University, Atlanta, Georgia 30303, USA}
\newcommand{\hanyang}{Hanyang University, Seoul 133-792, Korea}
\newcommand{\hiroshima}{Hiroshima University, Kagamiyama, Higashi-Hiroshima 739-8526, Japan}
\newcommand{\howard}{Department of Physics and Astronomy, Howard University, Washington, DC 20059, USA}
\newcommand{\ihepprot}{IHEP Protvino, State Research Center of Russian Federation, Institute for High Energy Physics, Protvino, 142281, Russia}
\newcommand{\illuiuc}{University of Illinois at Urbana-Champaign, Urbana, Illinois 61801, USA}
\newcommand{\inrras}{Institute for Nuclear Research of the Russian Academy of Sciences, prospekt 60-letiya Oktyabrya 7a, Moscow 117312, Russia}
\newcommand{\instpasczech}{Institute of Physics, Academy of Sciences of the Czech Republic, Na Slovance 2, 182 21 Prague 8, Czech Republic}
\newcommand{\isu}{Iowa State University, Ames, Iowa 50011, USA}
\newcommand{\jaea}{Advanced Science Research Center, Japan Atomic Energy Agency, 2-4 Shirakata Shirane, Tokai-mura, Naka-gun, Ibaraki-ken 319-1195, Japan}
\newcommand{\jyvaskyla}{Helsinki Institute of Physics and University of Jyv{\"a}skyl{\"a}, P.O.Box 35, FI-40014 Jyv{\"a}skyl{\"a}, Finland}
\newcommand{\karoly}{K\'aroly R\'oberts University College, H-3200 Gy\"ngy\"os, M\'atrai \'ut 36, Hungary}
\newcommand{\kek}{KEK, High Energy Accelerator Research Organization, Tsukuba, Ibaraki 305-0801, Japan}
\newcommand{\korea}{Korea University, Seoul, 136-701, Korea}
\newcommand{\kurchatov}{National Research Center ``Kurchatov Institute", Moscow, 123098 Russia}
\newcommand{\kyoto}{Kyoto University, Kyoto 606-8502, Japan}
\newcommand{\labllr}{Laboratoire Leprince-Ringuet, Ecole Polytechnique, CNRS-IN2P3, Route de Saclay, F-91128, Palaiseau, France}
\newcommand{\lahorelums}{Physics Department, Lahore University of Management Sciences, Lahore 54792, Pakistan}
\newcommand{\lawllnl}{Lawrence Livermore National Laboratory, Livermore, California 94550, USA}
\newcommand{\losalamos}{Los Alamos National Laboratory, Los Alamos, New Mexico 87545, USA}
\newcommand{\lund}{Department of Physics, Lund University, Box 118, SE-221 00 Lund, Sweden}
\newcommand{\maryland}{University of Maryland, College Park, Maryland 20742, USA}
\newcommand{\mass}{Department of Physics, University of Massachusetts, Amherst, Massachusetts 01003-9337, USA}
\newcommand{\michigan}{Department of Physics, University of Michigan, Ann Arbor, Michigan 48109-1040, USA}
\newcommand{\muhlenberg}{Muhlenberg College, Allentown, Pennsylvania 18104-5586, USA}
\newcommand{\myongji}{Myongji University, Yongin, Kyonggido 449-728, Korea}
\newcommand{\nagasaki}{Nagasaki Institute of Applied Science, Nagasaki-shi, Nagasaki 851-0193, Japan}
\newcommand{\nara}{Nara Women's University, Kita-uoya Nishi-machi Nara 630-8506, Japan}
\newcommand{\natmephi}{National Research Nuclear University, MEPhI, Moscow Engineering Physics Institute, Moscow, 115409, Russia}
\newcommand{\newmex}{University of New Mexico, Albuquerque, New Mexico 87131, USA}
\newcommand{\nmsu}{New Mexico State University, Las Cruces, New Mexico 88003, USA}
\newcommand{\ohio}{Department of Physics and Astronomy, Ohio University, Athens, Ohio 45701, USA}
\newcommand{\ornl}{Oak Ridge National Laboratory, Oak Ridge, Tennessee 37831, USA}
\newcommand{\orsay}{IPN-Orsay, Univ.~Paris-Sud, CNRS/IN2P3, Universit\'e Paris-Saclay, BP1, F-91406, Orsay, France}
\newcommand{\peking}{Peking University, Beijing 100871, P.~R.~China}
\newcommand{\pnpi}{PNPI, Petersburg Nuclear Physics Institute, Gatchina, Leningrad region, 188300, Russia}
\newcommand{\riken}{RIKEN Nishina Center for Accelerator-Based Science, Wako, Saitama 351-0198, Japan}
\newcommand{\rikjrbrc}{RIKEN BNL Research Center, Brookhaven National Laboratory, Upton, New York 11973-5000, USA}
\newcommand{\rikkyo}{Physics Department, Rikkyo University, 3-34-1 Nishi-Ikebukuro, Toshima, Tokyo 171-8501, Japan}
\newcommand{\saispbstu}{Saint Petersburg State Polytechnic University, St.~Petersburg, 195251 Russia}
\newcommand{\seoulnat}{Department of Physics and Astronomy, Seoul National University, Seoul 151-742, Korea}
\newcommand{\stonybrkc}{Chemistry Department, Stony Brook University, SUNY, Stony Brook, New York 11794-3400, USA}
\newcommand{\stonycrkp}{Department of Physics and Astronomy, Stony Brook University, SUNY, Stony Brook, New York 11794-3800, USA}
\newcommand{\sungskku}{Sungkyunkwan University, Suwon, 440-746, Korea}
\newcommand{\tenn}{University of Tennessee, Knoxville, Tennessee 37996, USA}
\newcommand{\titech}{Department of Physics, Tokyo Institute of Technology, Oh-okayama, Meguro, Tokyo 152-8551, Japan}
\newcommand{\tsukuba}{Center for Integrated Research in Fundamental Science and Engineering, University of Tsukuba, Tsukuba, Ibaraki 305, Japan}
\newcommand{\vandy}{Vanderbilt University, Nashville, Tennessee 37235, USA}
\newcommand{\weizmann}{Weizmann Institute, Rehovot 76100, Israel}
\newcommand{\wigner}{Institute for Particle and Nuclear Physics, Wigner Research Centre for Physics, Hungarian Academy of Sciences (Wigner RCP, RMKI) H-1525 Budapest 114, POBox 49, Budapest, Hungary}
\newcommand{\yonsei}{Yonsei University, IPAP, Seoul 120-749, Korea}
\newcommand{\zagreb}{University of Zagreb, Faculty of Science, Department of Physics, Bijeni\v{c}ka 32, HR-10002 Zagreb, Croatia}
\affiliation{\abilene}
\affiliation{\augie}
\affiliation{\banaras}
\affiliation{\barc}
\affiliation{\baruch}
\affiliation{\bnlcoll}
\affiliation{\bnlphys}
\affiliation{\caucr}
\affiliation{\charlesczech}
\affiliation{\chonbuk}
\affiliation{\ciae}
\affiliation{\cns}
\affiliation{\colorado}
\affiliation{\columbia}
\affiliation{\czechtech}
\affiliation{\elte}
\affiliation{\ewha}
\affiliation{\fsu}
\affiliation{\gsu}
\affiliation{\hanyang}
\affiliation{\hiroshima}
\affiliation{\howard}
\affiliation{\ihepprot}
\affiliation{\illuiuc}
\affiliation{\inrras}
\affiliation{\instpasczech}
\affiliation{\isu}
\affiliation{\jaea}
\affiliation{\jyvaskyla}
\affiliation{\karoly}
\affiliation{\kek}
\affiliation{\korea}
\affiliation{\kurchatov}
\affiliation{\kyoto}
\affiliation{\labllr}
\affiliation{\lahorelums}
\affiliation{\lawllnl}
\affiliation{\losalamos}
\affiliation{\lund}
\affiliation{\maryland}
\affiliation{\mass}
\affiliation{\michigan}
\affiliation{\muhlenberg}
\affiliation{\myongji}
\affiliation{\nagasaki}
\affiliation{\nara}
\affiliation{\natmephi}
\affiliation{\newmex}
\affiliation{\nmsu}
\affiliation{\ohio}
\affiliation{\ornl}
\affiliation{\orsay}
\affiliation{\peking}
\affiliation{\pnpi}
\affiliation{\riken}
\affiliation{\rikjrbrc}
\affiliation{\rikkyo}
\affiliation{\saispbstu}
\affiliation{\seoulnat}
\affiliation{\stonybrkc}
\affiliation{\stonycrkp}
\affiliation{\sungskku}
\affiliation{\tenn}
\affiliation{\titech}
\affiliation{\tsukuba}
\affiliation{\vandy}
\affiliation{\weizmann}
\affiliation{\wigner}
\affiliation{\yonsei}
\affiliation{\zagreb}
\author{A.~Adare} \affiliation{\colorado} 
\author{C.~Aidala} \affiliation{\losalamos} \affiliation{\michigan} 
\author{N.N.~Ajitanand} \affiliation{\stonybrkc} 
\author{Y.~Akiba} \affiliation{\riken} \affiliation{\rikjrbrc} 
\author{R.~Akimoto} \affiliation{\cns} 
\author{J.~Alexander} \affiliation{\stonybrkc} 
\author{M.~Alfred} \affiliation{\howard} 
\author{K.~Aoki} \affiliation{\kek} \affiliation{\riken} 
\author{N.~Apadula} \affiliation{\isu} \affiliation{\stonycrkp} 
\author{H.~Asano} \affiliation{\kyoto} \affiliation{\riken} 
% ...
\collaboration{PHENIX Collaboration} \noaffiliation

\date{\today}


\begin{abstract}


Measurements of anisotropic flow Fourier coefficients ($v_n$) for 
inclusive charged particles and identified hadrons $\pi^{\pm}$, $K^{\pm}$, 
$p$, and $\bar{p}$ produced at midrapidity in Cu$+$Au collisions at 
$\sqrt{s_{_{NN}}}=200$ GeV are presented. The data were collected in 2012 
by the PHENIX experiment at the Relativistic Heavy Ion Collider (RHIC). 
The particle azimuthal distributions with respect to different order 
symmetry planes $\Psi_n$, for $n$~=~1, 2, and~3 are studied as a 
function of transverse momentum $p_T$ over a broad range of collision
centralities. Mass ordering, as expected from hydrodynamic flow, is 
observed for all three harmonics. The charged-particle results are 
compared to hydrodynamical and transport model calculations. We also 
compare these Cu$+$Au results with those in Cu$+$Cu and Au$+$Au collisions 
at the same $\sqrt{s_{_{NN}}}$, and find that the $v_2$ and $v_3$, as a 
function of transverse momentum, follow a common scaling with 
$1/(\varepsilon_n N_{\rm part}^{1/3})$.

\end{abstract}


\maketitle

		\section{Introduction}
		\label{sec:introduction}

Measurements of azimuthal anisotropies of particle emission in 
relativistic heavy ion collisions have proven to be an essential tool in 
probing the properties of the quark gluon plasma (QGP) produced in such 
collisions. These anisotropies can be quantified~\cite{Voloshin:1994mz} by 
the coefficients $v_n$ in the Fourier expansion of the particle 
distributions with respect to symmetry planes of the same-order $\Psi_n$ 
that are determined on an event-by-event basis: $dN/d\phi \propto 1 + 
\sum_{n=1} 2 v_{n} \cos(n(\phi-\Psi_{n}))$, where $n$ is the order of the 
harmonic, $\phi$ is the azimuthal angle of particles of a given type, and 
$\Psi_n$ is the azimuthal angle of the $n^{\rm th}$-order symmetry plane.  
Measurements of the second harmonic, which indicates the strength of the 
``elliptic flow'', led to the conclusion that the QGP produced at RHIC 
behaves as a nearly inviscid 
fluid~\cite{Adcox:2004mh,Adams:2005dq,Back:2004je,Shuryak:2004cy,Gyulassy:2004zy}. 
In the last decade, significant effort, both experimentally and 
theoretically, has gone towards quantifying the specific viscosity 
$\eta/s$ (shear viscosity over entropy density) of the produced QGP, as 
well as its temperature dependence.

Elliptic flow is thought to arise from the initial spatial anisotropy in 
the nuclear overlap zone, which has a lenticular shape in off-center 
nucleus-nucleus ($A$$+$$A$) collisions. This spatial anisotropy is then 
converted to a momentum-space anisotropy through the pressure gradients in 
the expanding fluid. Measurements of $v_2$ have been performed in 
symmetric $A$$+$$A$ collision for a variety of collision energies and particle 
species as a function of transverse momentum, rapidity, and system 
size~\cite{Barrette:1996rs,Alt:2003ab,Adler:2002pu,Adamczyk:2013gw,Alver:2006wh,Adare:2014bga,Aamodt:2010pa,ATLAS:2011ah,Chatrchyan:2012ta}. 
Various scaling properties have been explored with the goal of 
understanding the onset of QGP formation with center-of-mass energy and 
how its properties may vary. The elliptic flow scaled by the corresponding 
initial spatial eccentricity ($\varepsilon_{2}$) was found to follow a universal trend when plotted against the produced particle density in the 
transverse plane~\cite{Alt:2003ab,Adler:2002pu,Chatrchyan:2012ta} over a 
broad range of center-of-mass energies. In a more recent 
study~\cite{Adare:2014bga}, PHENIX showed that the transverse particle 
density is proportional to the third root of the number of participant 
nucleons $N_{\rm part}^{1/3}$ and that scaling with $(\varepsilon_2 N_{\rm 
part}^{1/3})$ removes the remaining system-size dependencies at various 
center-of-mass energies.

The first-order Fourier coefficient $v_1$, which is a measure of the 
strength of the ``directed flow'', has also been studied in symmetric 
$A$$+$$A$ collisions over a broad range of 
energies~\cite{Barrette:1996rs,Alt:2003ab, Back:2005pc, 
Adamczyk:2014ipa,Abelev:2013cva}. Most studies focus on measurements of 
\pt-integrated values of $v_1$ as a function of rapidity or 
pseudorapidity, and the slope of $dv_1/dy$ at midrapidity, which may yield 
information on the location of a first-order phase transition in the phase 
diagram of nuclear matter~\cite{Rischke:1987xe}. In symmetric $A$$+$$A$ 
collisions, if the nuclei are considered to be smooth spheres, $v_1$ is an 
odd function with respect to (pseudo)rapidity and vanishes at midrapidity, 
which is consistent with the \pt-integrated measurements.

Indeed, when the nuclei are taken as smooth spheres, all odd harmonics 
should vanish at midrapidity. However, event-by-event fluctuations in the 
initial geometry can lead to nonzero odd harmonics at 
midrapidity~\cite{Alver:2010gr}.  Sizable values for these harmonics have 
been measured at both RHIC 
($v_3$)~\cite{Adare:2014kci,Adare:2011tg,Adamczyk:2013waa} and the 
Large Hadron Collider 
($v_3$ and $v_5)$~\cite{ALICE:2011ab,ATLAS:2012at,Chatrchyan:2013kba}. 
Evidence for a small rapidity-even component of $v_1$ at midrapidity has 
also been observed~\cite{Abelev:2013cva}. The combined experimental 
information from odd and even flow harmonics provides much more stringent 
constraints on the theoretical 
models~\cite{Schenke:2011bn,Gavin:2011gr,Han:2011iy,Qin:2010pf,Qiu:2011iv,Staig:2011wj} 
and the extracted QGP properties than measurements of elliptic flow alone.

Despite the wealth of experimental data and theoretical studies, 
uncertainties in the energy density deposition in the initial state of the 
heavy ion collisions remain a limiting factor in deducing the specific 
viscosity of the QGP. Asymmetric collision systems, such as Cu$+$Au, 
provide opportunities to study the effect of the initial geometry 
on the collective flow, particularly because odd harmonics may be enhanced 
at midrapidity beyond the fluctuation effects. 

In this paper, we present measurements of $v_1$, $v_2$, and $v_3$ of 
charged particles and identified hadrons $\pi^{\pm}$, $K^{\pm}$, $p$, and 
$\bar{p}$ produced at midrapidity in Cu$+$Au collisions at \sqsn=~200~GeV. 
In Sec.~\ref{sec:experimental} we present the experimental details of the 
measurements, and the sources of systematic uncertainties.  The results of 
the measurements are presented in Sec.~\ref{sec:results}.  In 
Sec.~\ref{sec:systemSize} we compare the flow results obtained in 
different collision systems and explore their scaling behavior, and in 
Sec.~\ref{sec:theory} we present comparisons to theoretical calculations.  
Sec.~\ref{sec:summary} summarizes our findings.

 \section{Experimental Details}
\label{sec:experimental}

\begin{figure}[htbp]
\includegraphics[width=0.96\linewidth]{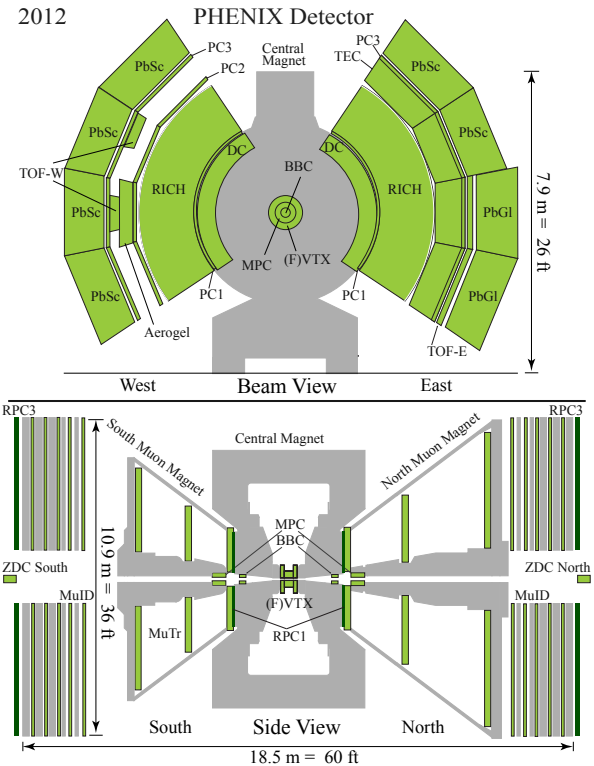}
\caption{
The PHENIX detector configuration for RHIC Run-12 data taking period}
\label{Fig:PHENIX}
\end{figure}

The PHENIX experiment is designed for the study of nuclear matter in 
extreme conditions using a wide variety of experimental observables. The 
detector, optimized for the high-multiplicity environment of 
ultra-relativistic heavy ion collisions, comprises two central-arm 
spectrometers (East and West), two muon spectrometers (at forward and 
backward rapidity), and a set of detectors used to determine the global 
properties of the collisions.  Figure~\ref{Fig:PHENIX} shows a schematic 
diagram of the PHENIX detector for the data recorded in 2012. The upper 
drawing shows a beam-axis view of the two central spectrometer arms, 
covering the pseudorapidity region $|\eta|<$ 0.35. The lower drawing shows 
a side view of the two forward-rapidity muon arms (North and South) and 
the global detectors. A detailed description of the complete set of 
detectors is given in Ref.~\cite{Adcox:2003zm}.

The analysis presented here employs the global detectors, drift chamber 
(DC), three layers of multi-wire proportional chambers (PC1, PC2, and 
PC3), the time-of-flight detectors (TOFE, TOFW), and the electromagnetic 
calorimeter (EMCal).  The global system includes the beam-beam counters 
(BBCs), zero degree calorimeters (ZDCs) and the shower maximum detectors 
(SMDs).  Below, we give a brief description of each of these detector 
sub-systems and their role in the present analysis.

	\subsection{Global Detectors}
	\label{s:global}

The BBCs are located at $\pm$144~cm from the nominal interaction point 
along the beam line, cover $2\pi$ in azimuth, and span the pseudorapidity 
range 3.0~$<|\eta|<$~3.9. Each BBC comprises 64 \Cher telescopes, arranged 
radially around the beam line.  The BBCs provide the main interaction 
trigger for the experiment and are also used in the determination of the 
collision vertex position along the beam axis ($z$-vertex) with $\sigma_z$ 
= 0.6 cm resolution and the centrality of the collisions.  The event 
centrality class in Cu$+$Au collisions is determined as a percentile of 
the total charge measured in the BBC from both sides.  The BBCs also 
provide the start time for the time-of-flight measurement with a timing 
resolution around $\sigma_t = 40$ ps in central Cu$+$Au 
collisions~\cite{Adcox:2003zm}.

The ZDCs~\cite{Adler:2003sp} are hadronic calorimeters located forward and 
backward of the PHENIX detector, along the beam line. Each ZDC is 
subdivided into three identical modules of two interaction lengths. They 
cover a pseudorapidity range of $|\eta|>6.5$ and measure the energy of 
spectator neutrons with an energy resolution of 
$\sigma(E)/E=85\%/\sqrt{E}+9.1\%$.  The SMDs~\cite{Adler:2003sp} are 
scintillator strip hodoscopes located 
between the first and second ZDC modules, a location corresponding 
approximately to the maximum of the hadronic shower. The horizontal 
coordinate is sampled by seven scintillator strips of 15 mm width, while 
the vertical coordinate is sampled by eight strips of 20 mm width. The 
active area of each SMD is 105 mm $\times$ 110 mm (horizontal $\times$ 
vertical dimension). Scintillation light is delivered to a multichannel 
Hamamatsu PMT R5900-M16 by wavelength shifting fibers~\cite{Adler:2003sp}. 
A typical position resolution for SMD is $\sim$~0.1--0.3~cm.

	\subsection{Tracking and Particle Identification Detectors}
	\label{s:detectors}

The charged-particle momentum is reconstructed using the tracking system. 
This system comprises the DC, located outside an 
axially-symmetric magnetic field at a radial distance between 2.0~m and 
2.4~m, followed by PC1-3. The pattern recognition in the DC is based on a 
combinatorial Hough transform~\cite{CHT} in the track bend plane. A track 
model based on a field-integral look-up table determines the 
charged-particle momentum, the path length to the time-of-flight 
detector, and a projection of the track to the outer detectors.

\begin{figure*}[htbp]
\includegraphics[trim=0 95 0 0, clip=true, width=0.9\linewidth]{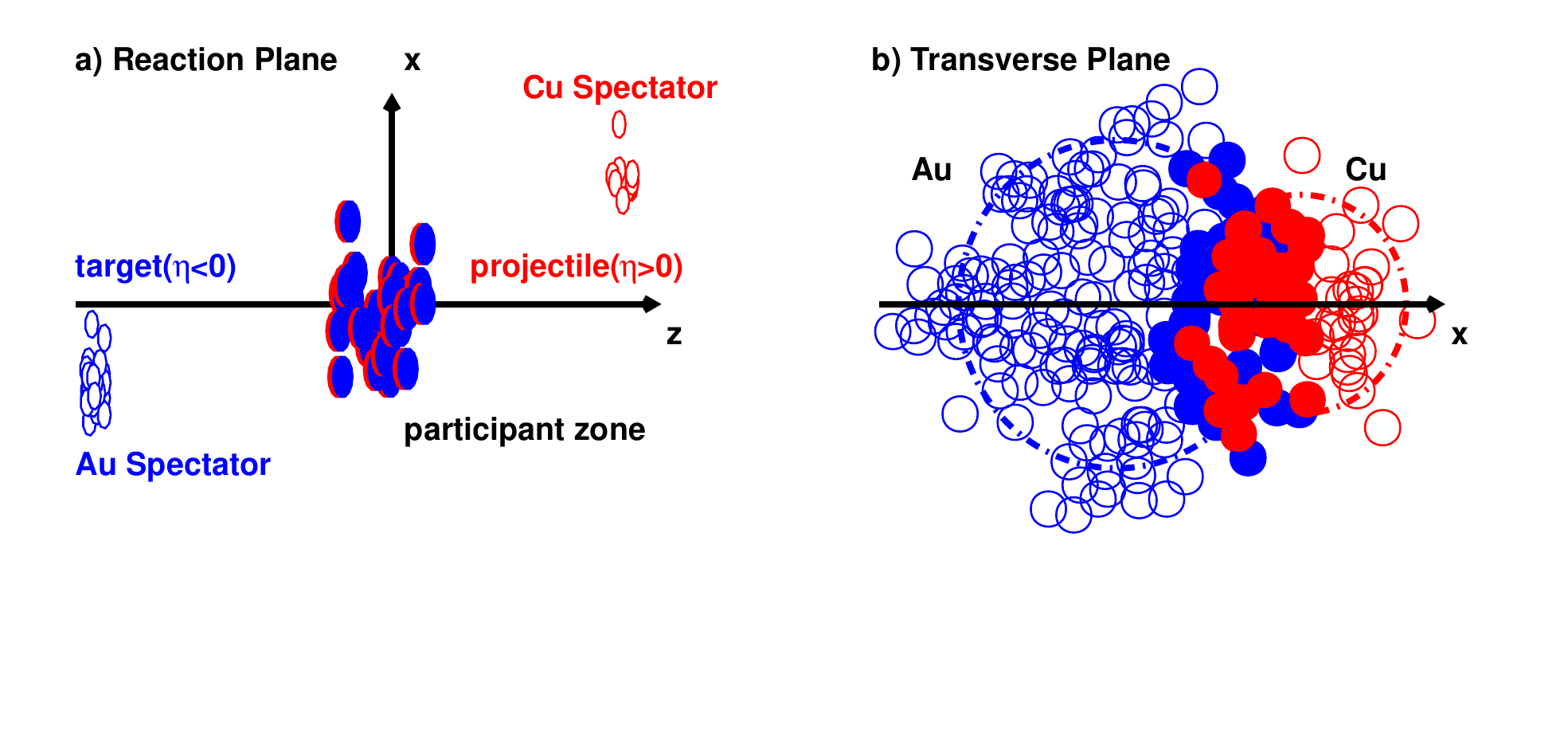}
\caption{
Sketch of a noncentral heavy-ion collision. See text for description of 
the figure.}
\label{fig:CuAu}
\end{figure*}

\begin{figure*}[htbp]
\includegraphics[width=1.0\linewidth]{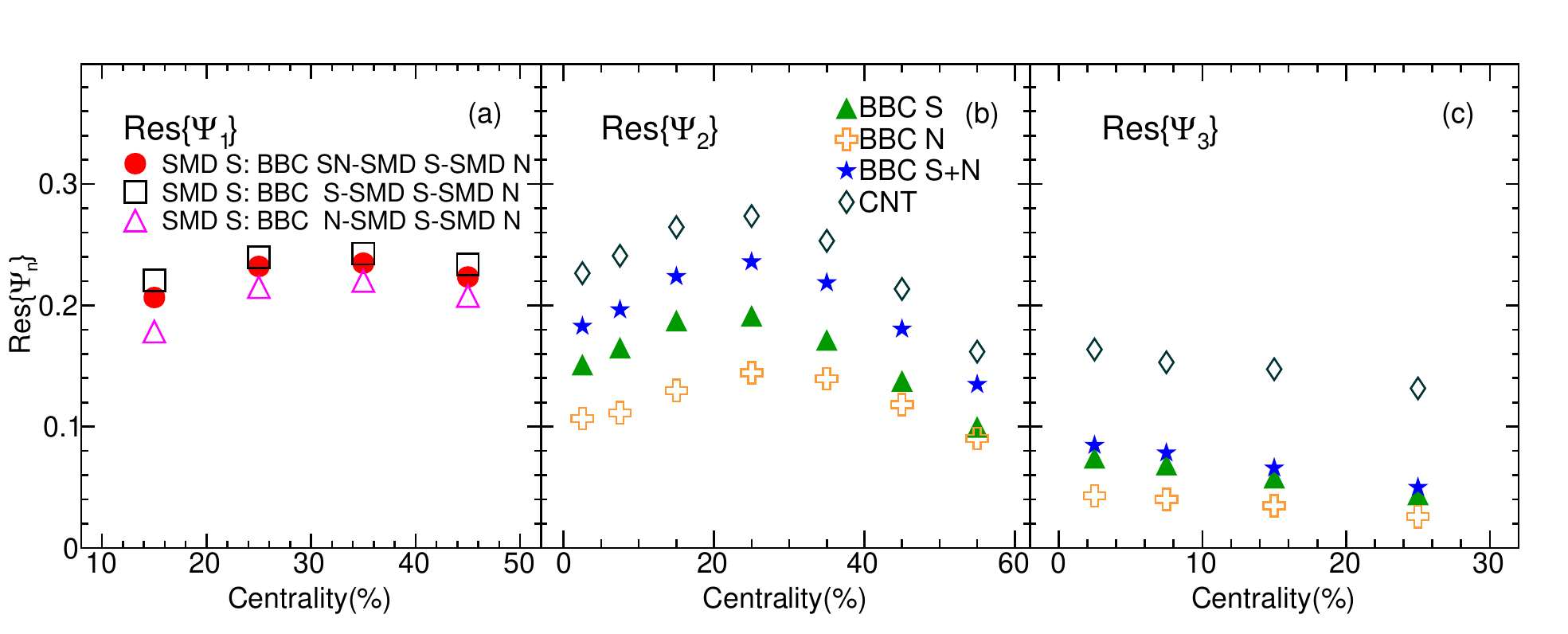}
\caption{
Panel (a) event-plane resolution as a function of centrality for the SMDS 
detectors. Panel (b) and (c) second- and third-order event-plane 
resolution. The BBC event-plane resolution is obtained from two sub-events 
and BBCS, BBCN, CNT from three sub-events as a function of 
centrality.}\label{fig:ev-reso}
\end{figure*}

The tracks are matched to hits registered in the PC3 and the EMCal, thus 
reducing the contribution of tracks originating from decays and 
$\gamma$-conversions.

The primary particle identification detectors used in this analysis are 
the time-of-flight detectors.  The different detectors in the east and 
west arms, use different technologies (scintillators and MRPCs 
respectively) and have different time 
resolutions~\cite{Adare:2013esx,Aizawa:2003zq}.  The total timing 
resolutions (including the start time measurement from the BBC) are 130~ps 
and 95~ps for east and west, respectively. Pion, kaon, and (anti)proton 
tracks are identified with over 97\% purity for 
$p_T<$~2~GeV/$c$~\cite{Adare:2013esx,Adare:2012vq} in both systems. For \pt between 
2--3~GeV/$c$, the purity of pions and protons is about 95\% and that of 
kaons is around 90\%.

	\subsection{Anisotropic Flow Measurement Technique}
	\label{sec:flowmethod}

The present measurements use the event-plane 
method~\cite{Poskanzer:1998yz} to quantify the azimuthal anisotropies of 
the particles produced in Cu$+$Au collisions. The $v_1$, $v_2$, and $v_3$ 
Fourier coefficients are determined as a function of centrality and \pt 
for inclusive charged particles and identified hadrons $\pi^{\pm}$, 
$K^{\pm}$, $p$, and $\bar{p}$ (with charge signs combined).

In the event-plane method, a measured event-plane direction $\Psi^{\rm 
obs}_n$ is determined for every event and for each order $n$.  The 
harmonic coefficients $v_n\{\Psi_n\}~=~\mean{\cos{n(\phi-\Psi^{\rm 
obs}_n)}}/{\rm Res}\{\Psi_n\}$ are then measured with respect to the event 
plane for each harmonic, where $\phi$ is the azimuthal angle of the hadron 
and ${\rm Res}\{\Psi_n\}$ is the event-plane resolution.

The collision geometry of a Cu$+$Au collision is shown in 
Fig.~\ref{fig:CuAu}(a) projected onto the reaction plane, and in 
Fig.~\ref{fig:CuAu}(b) projected onto the plane perpendicular to the beam 
axis. Figure~\ref{fig:CuAu}(a) shows direction of the projectile (Cu) and 
target (Au) spectators, which are bent away from the participant zone. 
There is an alternative picture, in which the spectators are attracted 
towards the center of the system, as discussed in~\cite{Alvioli:2010yk}. 
In this paper, we assume the former picture for the determination of the 
direction of the event-plane angle from the spectators.

As shown in Fig.~\ref{fig:CuAu}(a), the Cu spectators fly along the the 
positive-rapidity direction (North) and the Au spectators go towards the 
negative-rapidity direction (South).  The central position of the Au 
spectators is measured by the South SMD (SMDS) to determine the spectator 
plane $\Psi^{\rm SMDS}_1$.  The $v_1$ of charged and identified hadrons is 
measured with respect to $\Psi^{\rm SMDS}_1$, as indicated in 
Eq.~\ref{eq:v1}. Measurement with respect to the spectator plane is 
preferred over the first order event-plane determined by the distribution 
of the produced particles, because the distribution of the spectators is less 
distorted by momentum conservation effects.

\begin{equation}
v_{1} = -\langle\cos(\phi_{\rm track}-\Psi^{\rm SMDS}_{1})\rangle/{\rm Res}
(\Psi_{1}^{\rm SMDS})
\label{eq:v1}
\end{equation}

\begin{widetext}

\begin{eqnarray}
{\rm Res}(\Psi_{1}^{\rm SMDS})
&=&\langle\cos\left(\Psi_{1}^{\rm SMDS}-\Psi_{1}\right)\rangle \nonumber \\
&=& \sqrt{\frac{\langle\cos \left(\Psi_{1}^{\rm SMDS}
  -\Psi_{1}^{\rm SMDN}\right)\rangle
  \langle\cos \left(\Psi_{1}^{\rm SMDS}-\Psi_{1}^{\rm BBCSN}\right)\rangle} 
{\langle\cos\left(\Psi_{1}^{\rm SMDN}-\Psi_{1}^{\rm BBCSN}\right)\rangle}}
\label{eq:rp_res_3sub}
\end{eqnarray}

\end{widetext}

There is a negative sign in Eq.~\ref{eq:v1} to keep the convention in 
which the direction of projectile (Cu) spectators is positive. 
In Eq.~\ref{eq:rp_res_3sub} the 
resolution of $\Psi_1^{\rm SMDS}$ is calculated in 10\% centrality intervals with the three subevent 
method~\cite{Poskanzer:1998yz,Afanasiev:2009wq} by combining the other Cu 
spectator plane from the North SMD (SMDN) and the first order participant 
event-plane measured by the combined South and North BBCs (BBCSN).  
However, this method for determining the resolution assumes a
nonfluctuating nuclear-matter distribution.
Event-by-event fluctuations in the initial energy 
density of the collision will cause the $v_1$ signal to be different with 
respect to $\Psi_{1}^{\rm SMDS}$ and $\Psi_{1}^{\rm SMDN}$ due to the 
rapidity-symmetric component in the direct 
flow~\cite{PhysRevLett.111.232302}. To cover this uncertainty, the 
resolution of $\Psi_{1}^{\rm SMDS}$ is also calculated using the participant 
plane from either BBCS or BBCN and the differences are assigned as a 
systematic uncertainty.

The second ($\Psi_2$) and third ($\Psi_3$) order event planes are measured 
by the combination of BBCS and BBCN.  To determine the second and third 
order event-plane resolution from the BBC, we first measure the second and 
third order event planes with the BBCS (Au-going side), BBCN 
(Cu-going side) and central arm tracks (CNT).  The central-arm 
tracks are restricted to low \pt (0.2 $<p_T<$ 2.0~GeV/$c$) to minimize the 
contribution from jet fragments. The second and third order event-plane 
resolution of BBCS, BBCN, and CNT are calculated using three subevent 
methods with a combination of BBCS-BBCN-CNT. Then the second and third 
order event-plane resolutions of the BBC (including both BBCS and BBCN) 
are calculated with two subevent methods with a combination of BBC-CNT.

The event-plane resolutions for different subsystems are shown in 
Fig.~\ref{fig:ev-reso} as a function of centrality. Panel (a) of 
Fig.~\ref{fig:ev-reso} shows the resolution of the first-order event plane 
as measured by the SMDS using three different methods.  The first method uses a 
three subevent combination SMDS-BBCSN-SMDN, shown with circles, the 
second method shown with open squares uses a three subevent combination SMDS-BBCS-SMDN, and the third
method shown with open triangles uses the combination SMDS-BBCN-SMDN. 
The resolution of the second and third order event planes for BBC, BBCS, BBCN, 
and CNT are shown in panel (b) and (c) of Fig.~\ref{fig:ev-reso}, 
respectively.

	\subsection{Number of Participants and Eccentricity}
	\label{sec:Glauber}

A Monte-Carlo Glauber simulation was used to estimate the average 
number of participating nucleons $N_{\rm part}$ and the eccentricity

\begin{equation}
\varepsilon_n = \frac{\sqrt{\langle r^2\cos(n\phi) \rangle^2 + \langle r^2\sin (n\phi) \rangle^2}}{\langle r^2\rangle}
\label{eq:eccentricitydefinition}
\end{equation}

This simulation employed a Glauber model with a Woods-Saxon density 
profile and includes modeling of the BBC 
response~\cite{PhysRevLett.98.242302,ARNPS_57_205}.  The eccentricity 
defined in Eq.~\ref{eq:eccentricitydefinition} is 
also known as the participant eccentricity $\varepsilon_{\rm part}$ and 
includes the effect of fluctuations from the initial participant geometry. 
Table~\ref{tab:ecc_part_4systems} summarizes $N_{\rm part}$ and 
$\varepsilon_n$.

\begin{table*}[ht]
\caption{\label{tab:ecc_part_4systems} Number of participants and the 
participant eccentricity ($\eps_2$, $\eps_3$) from Monte-Carlo Glauber
calculations for Au$+$Au, Cu$+$Cu and Cu$+$Au collisions at 200 GeV}
\begin{ruledtabular}
\begin{tabular}{ccccccccc}
centrality &
\multicolumn{3}{c}{Au$+$Au 200 GeV} &
\multicolumn{2}{c}{Cu$+$Cu 200 GeV} &
\multicolumn{3}{c}{Cu$+$Au 200 GeV} \\
bin    & $N_{\rm part}$ & $\eps_2$ & $\eps_3$
       & $N_{\rm part}$ & $\eps_2$
       & $N_{\rm part}$ & $\eps_2$ & $\eps_3$ \\
\hline
 0\%--10\%  & 325.2     &  0.103 & 0.087 & 98.2 & 0.163 & 177.2 & 0.138 & 0.130 \\
            & $\pm$3.3  &$\pm$0.003 &$\pm$ 0.002&$\pm$2.4  &$\pm$0.003 &$\pm$5.2 &$\pm$0.011&$\pm$0.004 \\
10\%--20\%  & 234.6     &  0.200 & 0.122 & 73.6 & 0.241 & 132.4 & 0.204 & 0.161 \\
            &$\pm$4.7   &$\pm$0.005 &$\pm$ 0.004&$\pm$2.5  &$\pm$0.007 &$\pm$3.7 &$\pm$0.008&$\pm$0.005 \\
20\%--30\%  & 166.6     &  0.284 & 0.156 & 53.0 & 0.317 & 95.1  & 0.280 & 0.208 \\
            &$\pm$5.4   &$\pm$0.006 &$\pm$ 0.005&$\pm$1.9  &$\pm$0.006 &$\pm$3.2 &$\pm$0.008&$\pm$0.007 \\
30\%--40\%  & 114.2     &  0.356 & 0.198 & 37.3 & 0.401 & 65.7  & 0.357 & 0.266 \\
            &$\pm$4.4   &$\pm$0.006 &$\pm$ 0.008&$\pm$1.6  &$\pm$0.008 &$\pm$3.4 &$\pm$0.010&$\pm$0.010 \\
40\%--50\%  & 74.4      &  0.422 & 0.253 & 25.4 & 0.484 & 43.3  & 0.436 & 0.332 \\
            &$\pm$3.8   &$\pm$0.006 &$\pm$ 0.011&$\pm$1.3  &$\pm$0.008 &$\pm$3.0 &$\pm$0.013&$\pm$0.013 \\
50\%--60\%  & 45.5      &  0.491 & 0.325 & 16.7 & 0.579 & 26.8  & 0.523 & 0.412 \\
            &$\pm$3.3   &$\pm$0.005 &$\pm$ 0.018&$\pm$0.9  &$\pm$0.008 &$\pm$2.6 &$\pm$0.019&$\pm$0.019 \\
\end{tabular}
\end{ruledtabular}
\end{table*}

\subsection{Data set}
\label{sec:data}

The measurements presented here use data from Cu$+$Au collisions at 
$\sqrt{s_{_{NN}}}~=~200$~GeV collected by the PHENIX experiment at RHIC in 
2012. Minimum bias events triggered by the BBC recorded within $\pm$ 30 cm 
from the nominal interaction point along the z-axis were used. The events 
were examined to ensure that stable performance is seen in the detectors 
used in the analysis, namely DC, PC3, TOF, BBC, and SMD. A total of 
$3.6\times10^{9}$ events were analyzed.

	\subsection{Systematic Uncertainties} 
	\label{sec:systematics}

\begin{table}[htbp]
\caption{\label{tb:sysv1}Systematic uncertainties in the $v_1$ measurements.}
\begin{ruledtabular}
\begin{tabular}{ccccc}
$v_1$  &   Uncertainty Sources  & 10\%--20\% & 40\%--50\% & Type \\ 
\hline
$v_1$  & Event-plane & 20\% & 12\% & C \\
&      Background(absolute value)                      &  5$\times10^{-4}$  & 5$\times10^{-4}$ & A \\
&      Acceptance (absolute value)                      &  3$\times10^{-3}$  & 2$\times10^{-3}$ & C \\
\end{tabular}
\end{ruledtabular}
\end{table}

\begin{table}[htbp]
\caption{\label{tb:sysv23}
Systematic uncertainties given in percent on the $v_2$ and $v_3$ measurements.}
\begin{ruledtabular}
\begin{tabular}{ccccc}
$v_n$(n=2,3)  &   Uncertainty Sources  & 0\%--10\% & 20\%--30\% & Type \\ 
\hline
$v_2$  &      Event-plane   & 3\% & 4\% & B\\
&      Background               & 2\% & 2\% & A\\
&      Acceptance               & 2\% & 3\% & C\\
\\
$v_3$  &      Event-plane   & 3\% & 7\%  & B\\
&      Background               & 2\% & 2\%  & A\\
&      Acceptance               & 8\% & 10\% & C\\
\end{tabular}
\end{ruledtabular}
\end{table}

\begin{table}[htbp]
\caption{\label{tb:syspid_v1}
Systematic uncertainties in the measured $v_1$ for identified particles.}
\begin{ruledtabular}
 \begin{tabular}{cccc}
      species               & $p_{T}\le$ 2 GeV/$c$ & $p_{T}\ge$ 2 GeV/$c$  & Type \\ 
\hline
      pion     (absolute value)              & 1$\times 10^{-3}$ & 2$\times 10^{-3}$ & A \\
      kaon    (absolute value)              & 1$\times 10^{-3}$ & 3$\times 10^{-3}$ & A\\
      proton  (absolute value)             & 1$\times 10^{-3}$ & 3$\times 10^{-3}$ & A\\
\end{tabular}
\end{ruledtabular}
 \end{table}

\begin{table}[htbp]
\caption{\label{tb:syspid}
Systematic uncertainties in percent on the measured $v_2$ and $v_3$ for identified particles.}
\begin{ruledtabular}
 \begin{tabular}{cccc}
      species               & $p_{T}\le$ 2 GeV/$c$ & $p_{T}\ge$ 2 GeV/$c$  & Type \\ 
\hline
      pion                  & 3\% & 5\% & A \\
      kaon                  & 3\% & 10\% & A\\
      proton                & 3\% & 5\% & A\\
\end{tabular}
\end{ruledtabular}
 \end{table}

Tables~\ref{tb:sysv1}--\ref{tb:syspid} summarize the systematic
uncertainties for the measurements of $v_1$, $v_2$,
and $v_3$ for inclusive and identified charged hadrons, which are categorized by the types:

\begin{itemize}
\item[A] point-to-point uncertainties uncorrelated between $p_T$ bins;

\item[B] $p_{T}$-correlated, all points move in a correlated manner, but
not by the same factor;

\item[C] an overall normalization error in which all points move by the
same multiplicative factor independent of $p_T$.
\end{itemize}

Contributions to the uncertainties are from the following sources:
\begin{enumerate}
\item event-plane resolution correction,
\item event plane as measured using different detectors,
\item $v_n$ from background tracks,
\item acceptance dependencies
\item PID purity.
\end{enumerate}

The uncertainties from measurements of the event planes using different 
detectors are found to only weakly depend on $p_T$. For the measurement of 
$v_1$, the uncertainties are obtained by comparing the $v_1$ as measured 
with SMDS with alternately BBCN or BBCS used for resolution. For $v_2$ and 
$v_3$, the uncertainties are obtained by comparing the $v_2$ and $v_3$ as 
measured by the BBCN and BBCS. For the $v_1$ measurement, for the 
10\%--20\% centrality class we find a 20\% systematic uncertainty 
independent of $p_T$. For the 40\%--50\% centrality class, we find a 12\%
systematic uncertainty. For $v_2$, the systematic uncertainty is less than 
3\% for the 0\%--10\% centrality range and increases to 4\% for the 
centrality range 50\%--60\%. For $v_3$, a 3\% systematic uncertainty is 
found for 0\%--10\% centrality, increasing to 7\% for the 20\%--30\%
centrality range.

Background tracks that are not removed by the tracking selections as 
described in Sec.I\hspace{-.1em}I may influence the measured $v_n$. They 
can arise from particle decays, $\gamma$-conversions, or false track 
reconstruction.  We estimate the tracking background contribution by 
varying the width of the track-matching window in PC3 and comparing the results with and 
without the EMCal matching cut. We find that the absolute uncertainty for $v_{1}$
 is  less than $5\times10^{-4}$. For $v_{2}$ and $v_{3}$, the change is less than 2\%.

Systematic uncertainties of acceptance were evaluated using different 
subsets of the detector such as DC and TOF in the east and west arms. 
Differences in the $v_n$ measured using different arms may be caused by 
different detector alignment and performance. Maximum differences of order 
 3\% and 10\% were found for $v_2$ and $v_3$ respectively.  
These uncertainties have centrality dependence and minimal $p_T$ dependence.
 For $v_1$, maximum absolute uncertainty of  $3\times10^{-3}$ is found.
 These uncertainties are detailed further in Tables \ref{tb:sysv1} and \ref{tb:sysv23}. 

An additional systematic uncertainty in $v_n$ resulting from hadron 
misidentification is based on the PID purity estimates from the TOF 
detectors as discussed in Sec.I\hspace{-.1em}I. Pion, kaon, and proton 
species purity is greater than 90\% and the differences between their 
corresponding $v_n$ is less than a factor of two. For $v_{2}$ and $v_{3}$,
 an additional uncertainty of 3\% (type A) attributable to contamination from other species is found for 
particles with $p_T<2$~GeV/$c$, 5\% for higher $p_{T}$ pions and protons, and 
10\% for higher $p_{T}$ kaons. In the measurements of $v_1$, a common absolute uncertainty of 1$\times 10^{-3}$ is found for the three particle species for $p_T<2$~GeV/$c$, and at higher $p_T$ the uncertainties are 
2$\times 10^{-3}$ for pions  and 3$\times 10^{-3}$ for kaons and protons, respectively. The uncertainties due to
particle identification are to be added in quadrature to the values listed in Tables~\ref{tb:sysv1} and \ref{tb:sysv23}.  

 		\section{Results and discussion}
		\label{sec:results}

	\subsection{Harmonic flow results from Cu$+$Au collisions}
	\label{sec:chargedResults}

\begin{figure*}[!hthb]
\includegraphics[width=0.95\linewidth]{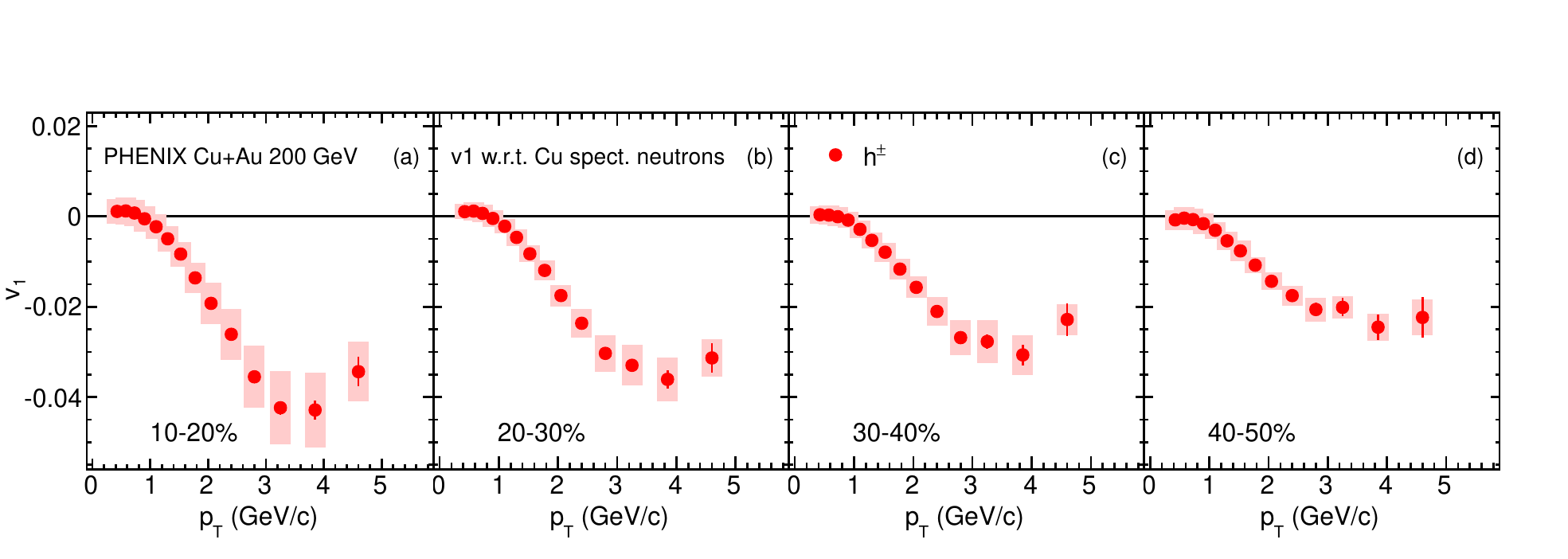}
\caption{
$v_1(p_T)$ for charged hadrons measured 
with respect to the Cu
spectator neutrons at midrapidity in Cu$+$Au collisions 
at $\sqrt{s_{_{NN}}}=200$~GeV.  
Error bars show the statistical uncertainties, 
and shaded boxes indicate the systematic uncertainties.}
\label{fig:v1}
\end{figure*}
\begin{figure*}[!hbt]
\includegraphics[width=0.9\linewidth]{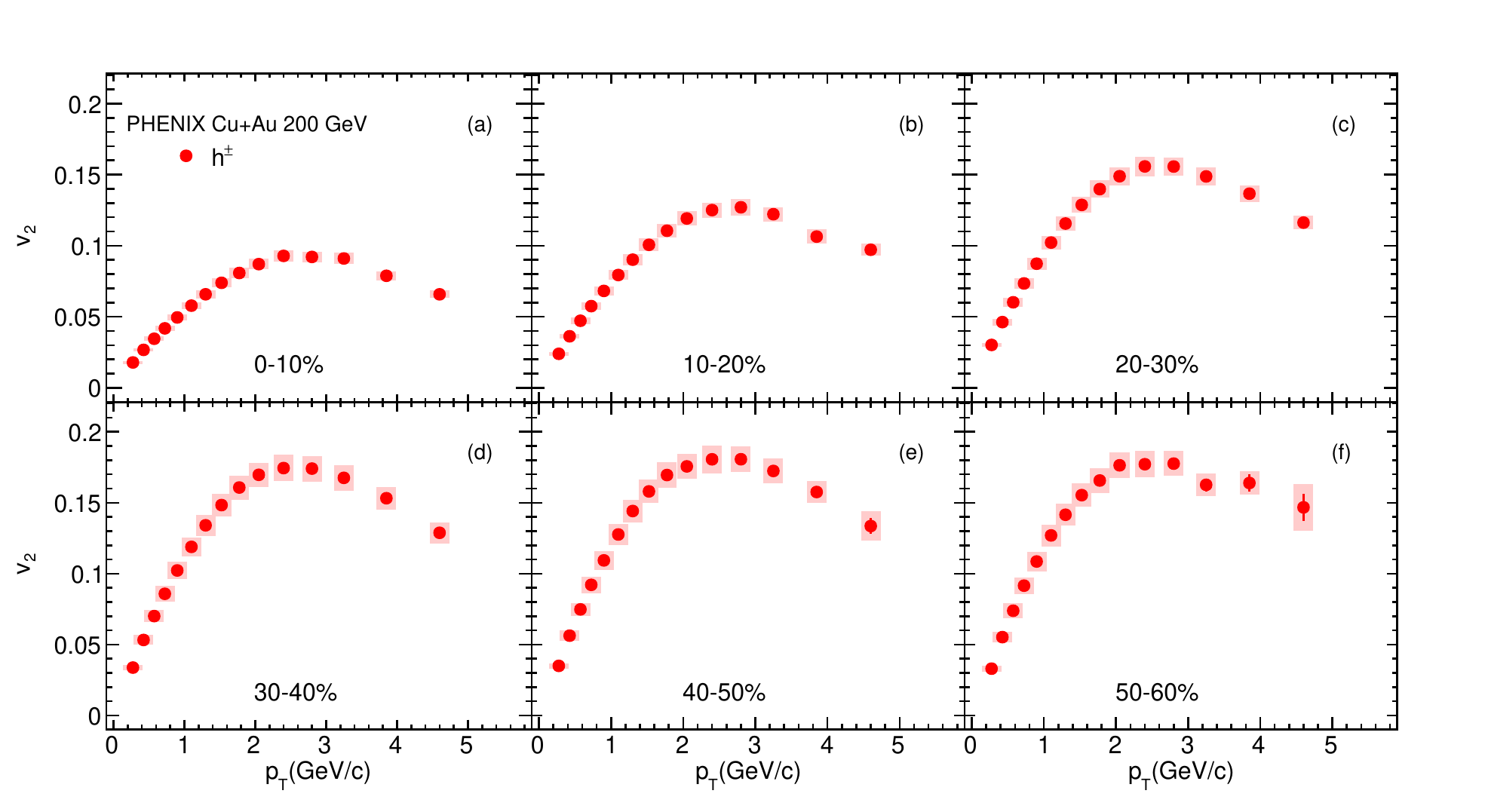}
\caption{ 
$v_2(p_T)$ for charged hadrons measured at midrapidity in Cu$+$Au 
collisions at $\sqrt{s_{_{NN}}}=200$~GeV.
Uncertainties are as in Fig.~\protect\ref{fig:v1}.
}
\label{fig:v2}
\end{figure*}
\begin{figure*}[!th]
\includegraphics[width=0.9\linewidth]{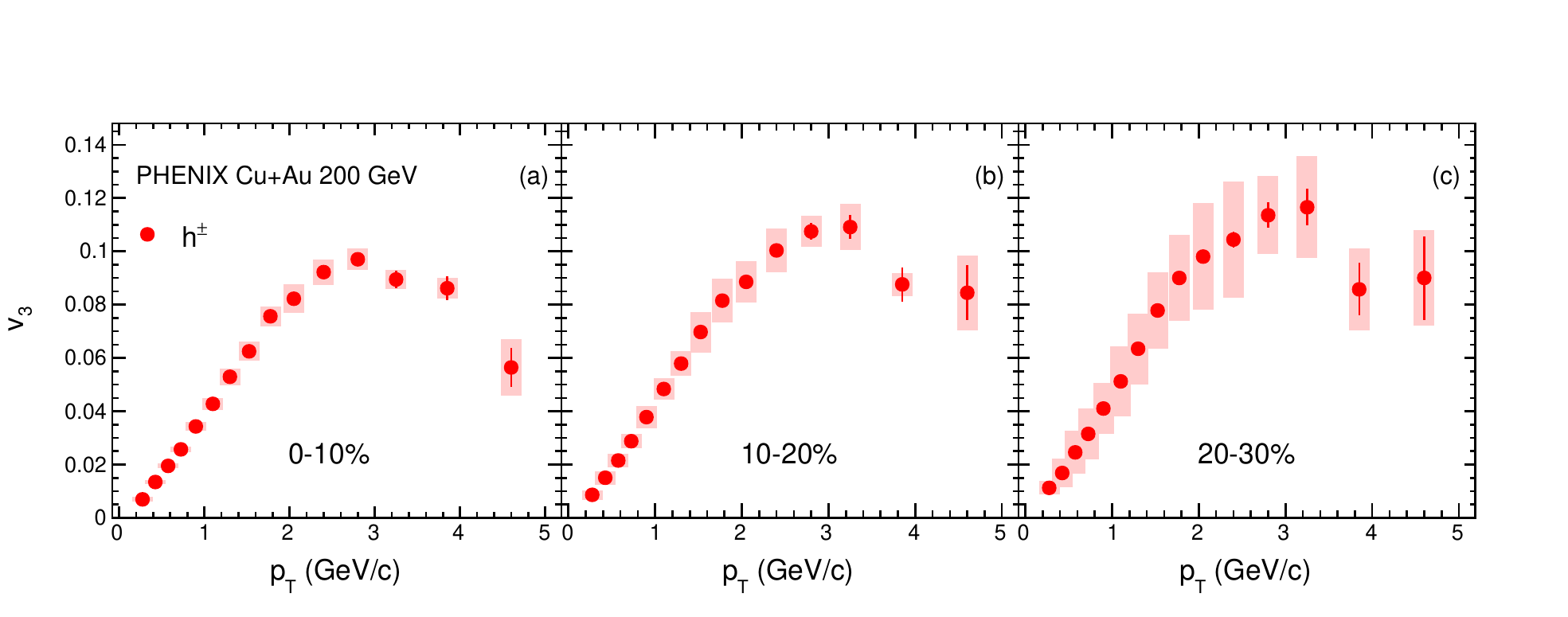}
\caption{ 
$v_3(p_T)$ for charged hadrons measured at midrapidity in Cu$+$Au 
collisions at $\sqrt{s_{_{NN}}}=200$ GeV. 
Uncertainties are as in Fig.~\protect\ref{fig:v1}.
}
\label{fig:v3}
\end{figure*}

Figures~\ref{fig:v1}--\ref{fig:v3} show the $v_1$, $v_2$, and $v_3$ 
results for charge-combined hadrons measured as a function of \pt in 
Cu$+$Au collisions at $\sqrt{s_{_{NN}}}$ = 200 GeV.  Different centrality 
intervals are studied. The filled circles show the $v_n(p_T)$ data, and 
the systematic uncertainties are shown with the shaded boxes.

The $v_1(p_T)$ measurements shown in Fig.~\ref{fig:v1} are performed with 
respect to the event plane determined by spectator neutrons from the Au 
nucleus.  To align with previous conventions, we flip the sign so that it 
is effectively with respect to the spectator neutrons from the Cu nucleus, 
as noted in Sec.~\ref{sec:flowmethod}.
In all centrality intervals, high $p_{T}$ particles 
at midrapidity move in the direction opposite of the Cu nucleus spectator 
neutrons, as indicated by the negative $v_{1}$ values. Low $p_{T}$ 
particles might then be expected to move in the opposite direction by 
conservation of momentum, and there is a hint of this effect though not 
beyond current systematic uncertainties. The $v_1$ component is consistent 
with zero for \pt$<1$ GeV/$c$ and its absolute value increases at higher 
\pt. The maximum of the absolute value decreases from central to 
peripheral collisions. This is contrary to the centrality dependence of 
$v_2$ where the values increase from the most central 0\%--10\% 
collisions, up to the 30\%--40\% centrality class. This trend in $v_2$ is 
expected from the initial geometry, because the ellipticity of 
the participant zone $\varepsilon_2$ (see Table 
~\ref{tab:ecc_part_4systems}) increases in the peripheral collisions. The 
$v_2(p_T)$ values in the 30\%--40\%, 40\%--50\%, and 50\%--60\% Cu$+$Au 
centrality classes, shown in Fig.~\ref{fig:v2} are consistent with each 
other, showing very little, if any centrality dependence. The $v_2$ and 
$v_3$ values are positive, as previously observed in symmetric collisions 
systems. For all three harmonics, the magnitude of the signal increases 
with \pt up to about \pt=~3~GeV/$c$, and then tends to decrease. This may 
indicate a change in the dominant production mechanism, e.g., an 
increasing contribution from jet fragments, or it may be due to the fact 
that higher \pt particles escape the fireball with fewer interactions.

The $v_3$ component (Fig.~\ref{fig:v3}) has weak centrality dependence, a 
behavior which is similar in symmetric $A$$+$$A$ 
collisions~\cite{Adare:2014kci,Adare:2011tg}, where the triangular flow at 
midrapidity is completely driven by the event-by-event fluctuations of the 
interaction zone. These fluctuations are also present in the asymmetric 
Cu$+$Au collisions and are expected to play a similar role. In 
Sec.~\ref{sec:systemSize} we compare the flow results obtained in 
different collisions systems and explore their scaling behavior.

	\subsection{Identified particle flow results}
	\label{sec:PIDflow}

\begin{figure*}[htbp]
\includegraphics[width=1.0\linewidth]{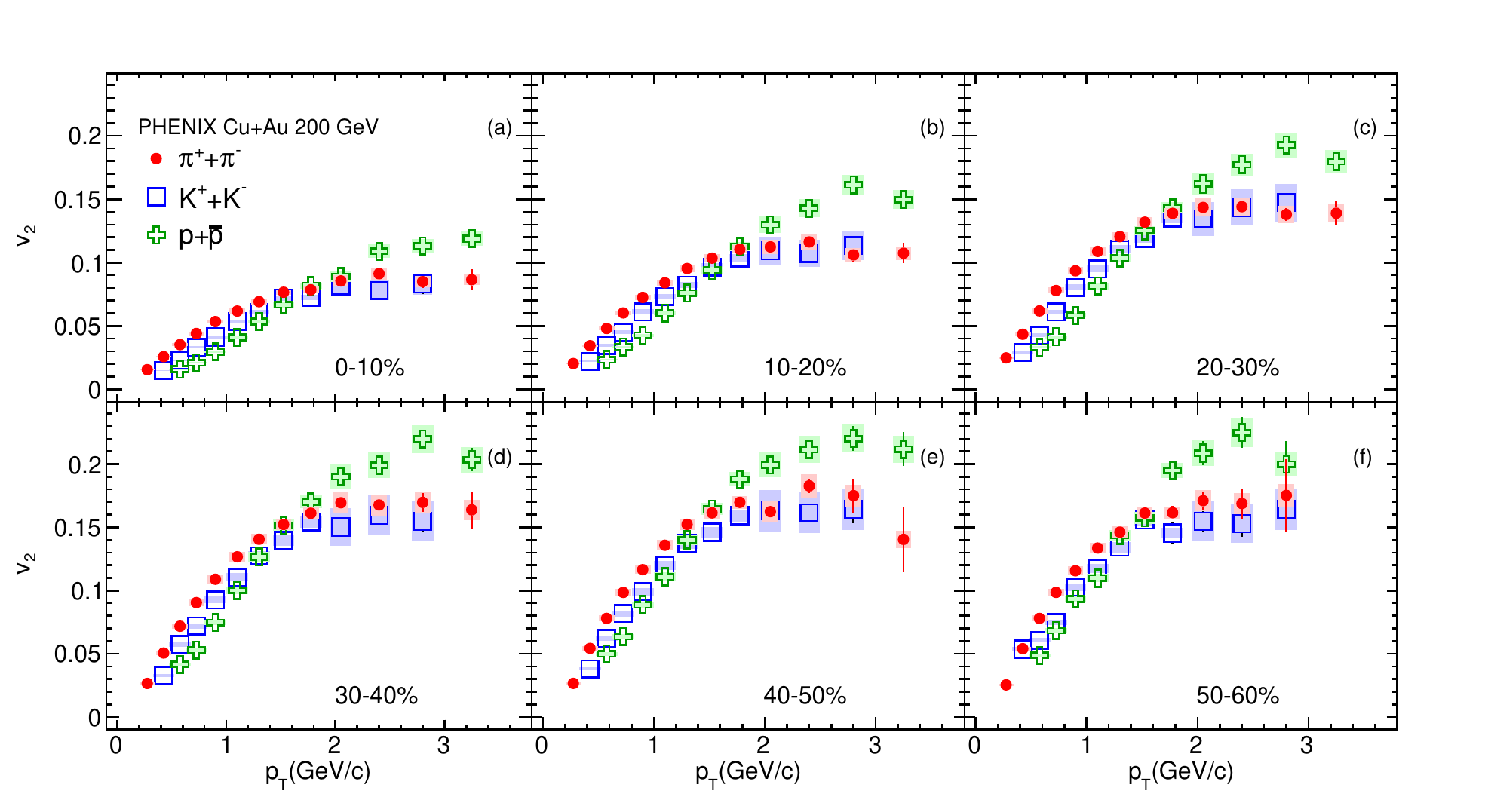}
\caption{
The second-order Fourier coefficients $v_2(p_T)$ for charge-combined 
identified hadrons $\pi^{\pm}$, $K^{\pm}$, $p$, and $\bar{p}$ measured at 
midrapidity in Cu$+$Au collisions at $\sqrt{s_{_{NN}}}$ = 200 GeV for the 
centrality classes marked in each panel. The symbols represent the 
measured $v_2(p_T)$ values, the error bars show the statistical 
uncertainties, and the shaded boxes indicate the systematic uncertainties 
from PID. The full systematic uncertainties, that are mostly common to all 
particle species are shown in Table~\ref{tb:sysv23}. }
\label{fig:v2PID}
\end{figure*}

\begin{figure*}[htbp]
\begin{minipage}{0.48\linewidth}
\includegraphics[width=1.0\linewidth]{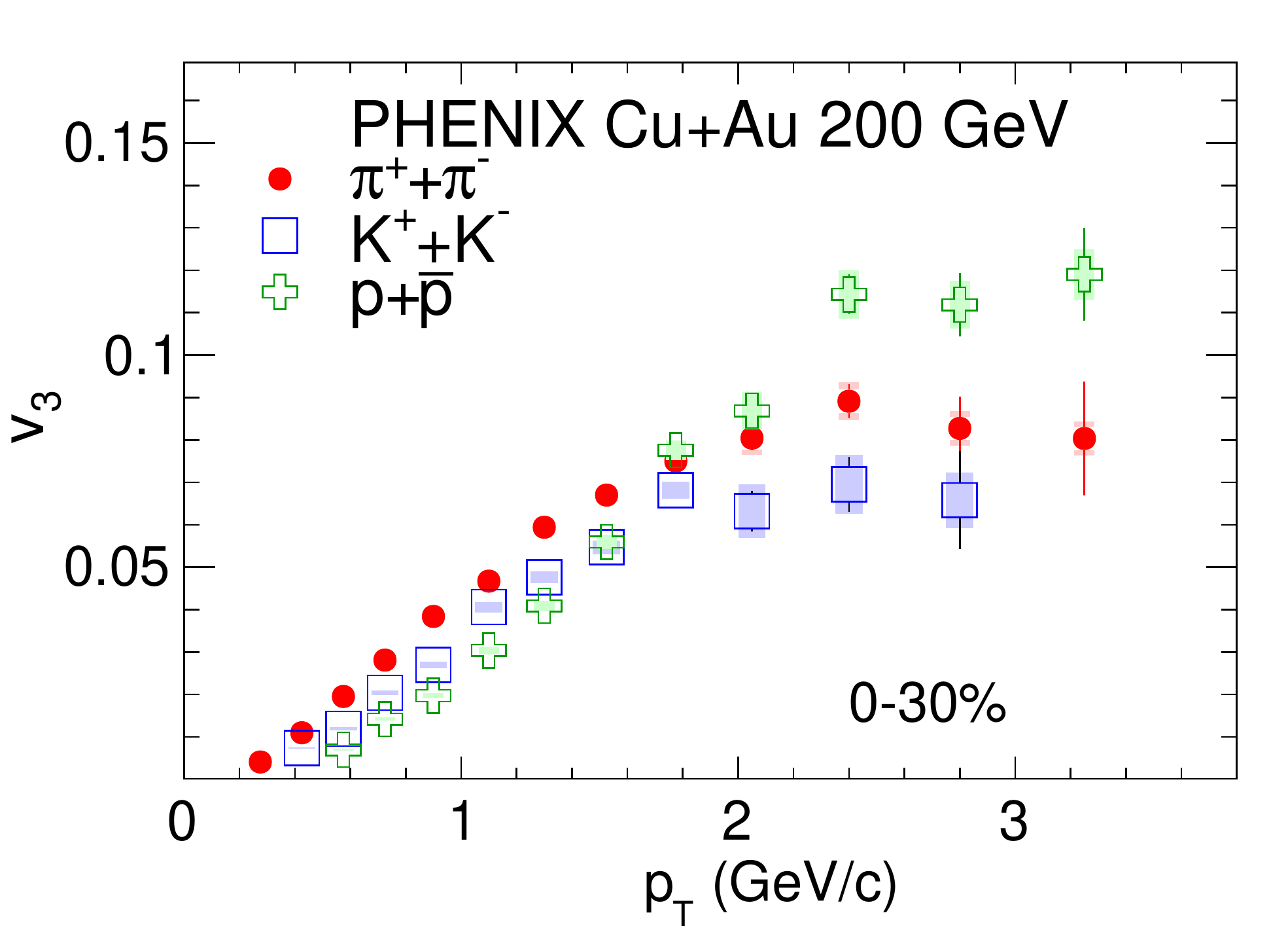}
\caption{
The third-order Fourier coefficients $v_3(p_T)$ for charge-combined 
identified hadrons $\pi^{\pm}$, $K^{\pm}$, $p$, and $\bar{p}$ measured at 
midrapidity in Cu$+$Au collisions at $\sqrt{s_{_{NN}}}$ = 200 GeV for 0\%--30\% 
centrality. The symbols represent the measured $v_3(p_T)$ values, the 
error bars show the statistical uncertainties, and the shaded boxes 
indicate the systematic uncertainties from PID. The full systematic 
uncertainties, that are mostly common to all particle species are shown in 
Table~\protect\ref{tb:sysv23}. }
\label{fig:v3PID}
\end{minipage}
\begin{minipage}{0.48\linewidth}
\hspace{0.3\linewidth}
\includegraphics[width=1.0\linewidth]{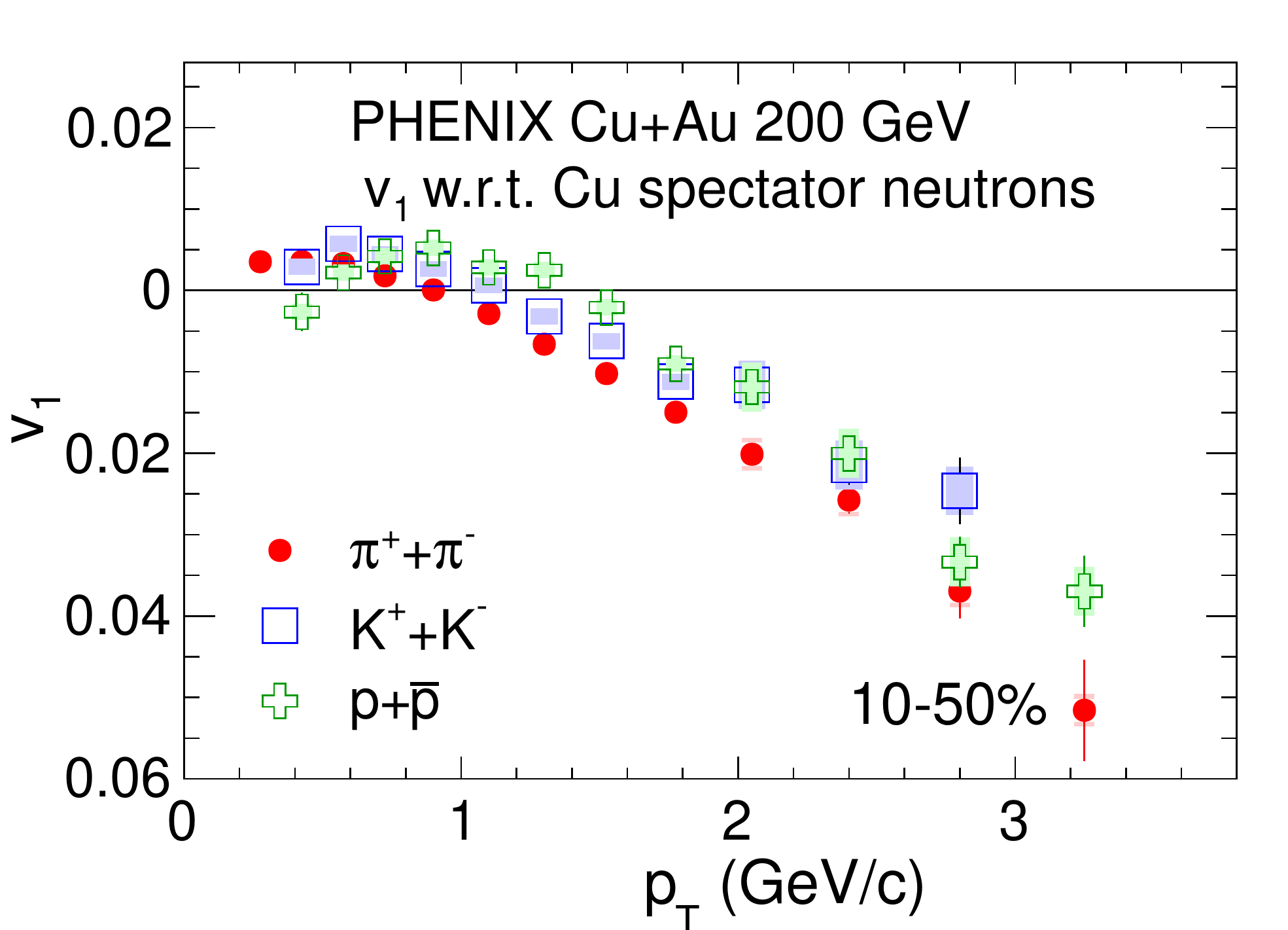}
\caption{
The first-order Fourier coefficients $v_1(p_T)$ for charge-combined 
identified hadrons $\pi^{\pm}$, $K^{\pm}$, $p$, and $\bar{p}$ measured at 
midrapidity in Cu$+$Au collisions at $\sqrt{s_{_{NN}}}$ = 200 GeV for 
10\%--50\% centrality. The symbols represent the measured $v_1(p_T)$ 
values with respect to the Cu spectator neutrons, the error bars show the 
statistical uncertainties, and the shaded boxes indicate the systematic 
uncertainties from PID. The full systematic uncertainties, that are mostly 
common to all particle species are shown in Table~\ref{tb:sysv1}.}
\label{fig:v1PID}
\end{minipage}
\end{figure*}

Figures~\ref{fig:v2PID} and~\ref{fig:v3PID} show the particle-species 
dependence of $v_2$ and $v_3$ in Cu$+$Au collisions.  Results are 
presented for charge-combined $\pi^{\pm}$, $K^{\pm}$, $p$, and $\bar{p}$. 
The measured $v_n(p_T)$ values are shown with points, and the shaded boxes 
represent the species-dependent type A systematic uncertainties. The type 
B and C systematic uncertainties shown in Table~\ref{tb:sysv23} are largely 
common for all particle species. For the odd harmonics, to improve the 
statistical significance of the results the measurements for identified 
particles are performed in a single centrality interval, namely 0\%--30\% 
for $v_3(p_T)$ and 10\%--50\% for $v_1(p_T)$.
  
There are two trends common to both $n=2,3$ results shown in 
Figs.~\ref{fig:v2PID} and~\ref{fig:v3PID}:  First, in the low-\pt region 
the anisotropy appears largest for the lightest hadron and smallest for 
the heaviest hadron.  A similar mass ordering is also predicted by 
hydrodynamics, in which all particles are moving in a common velocity 
field. Second, for \pt$\geq 2$~GeV/$c$ this mass dependence is reversed, 
such that the anisotropy is larger for the baryons than it is for mesons 
at the same \pt.  These patterns have been observed previously in $v_n$ 
measurements for identified particles in Au$+$Au collisions at RHIC. The 
$v_1(p_T)$ values, presented in Fig.~\ref{fig:v1PID}, also show mass 
ordering, although these measurements have larger overall systematic and 
statistical uncertainties than $v_2(p_T)$ and $v_3(p_T)$.  As in the case 
of $v_1(p_T)$ for charged particles described in Sec.~\ref{sec:results} 
(Fig.~\ref{fig:v1}), we note that although the values of $v_1(p_T)$ for 
each species appear to be positive at low \pt, if the full systematic 
uncertainty of type B and C is taken into account, a definitive conclusion 
can not be drawn about the overall sign of the bulk directed flow. The 
mass dependence in the collective flow at the low-\pt is a generic feature 
of hydrodynamical models. The dependence on valence quark number in the 
intermediate-\pt region has been associated with the development of flow 
in the partonic phase of the fireball evolution and subsequent 
hadronization by parton coalescence~\cite{Adare:2006ti}.

	\subsection{System size dependence}
	\label{sec:systemSize}

It is interesting to compare the charged-hadron $v_n(\pt)$ results for 
different collision systems measured in the same experiment at the same 
center-of-mass energy. PHENIX has previously studied anisotropic flow 
harmonics in symmetric Au$+$Au and Cu$+$Cu collisions at 
$\sqrt{s_{_{NN}}}$~=~200~GeV~\cite{Adare:2014bga,Adare:2011tg}. By varying 
the system size and the centrality selection, one can study the effects of 
the initial geometry on the observed flow coefficients. We will first 
compare the results obtained in different collision systems for the same 
centrality selections, and then explore possible scaling behaviors.
  
\begin{figure*}[htbp]
\begin{minipage}{1.0\linewidth}
\includegraphics[width=1.0\linewidth]{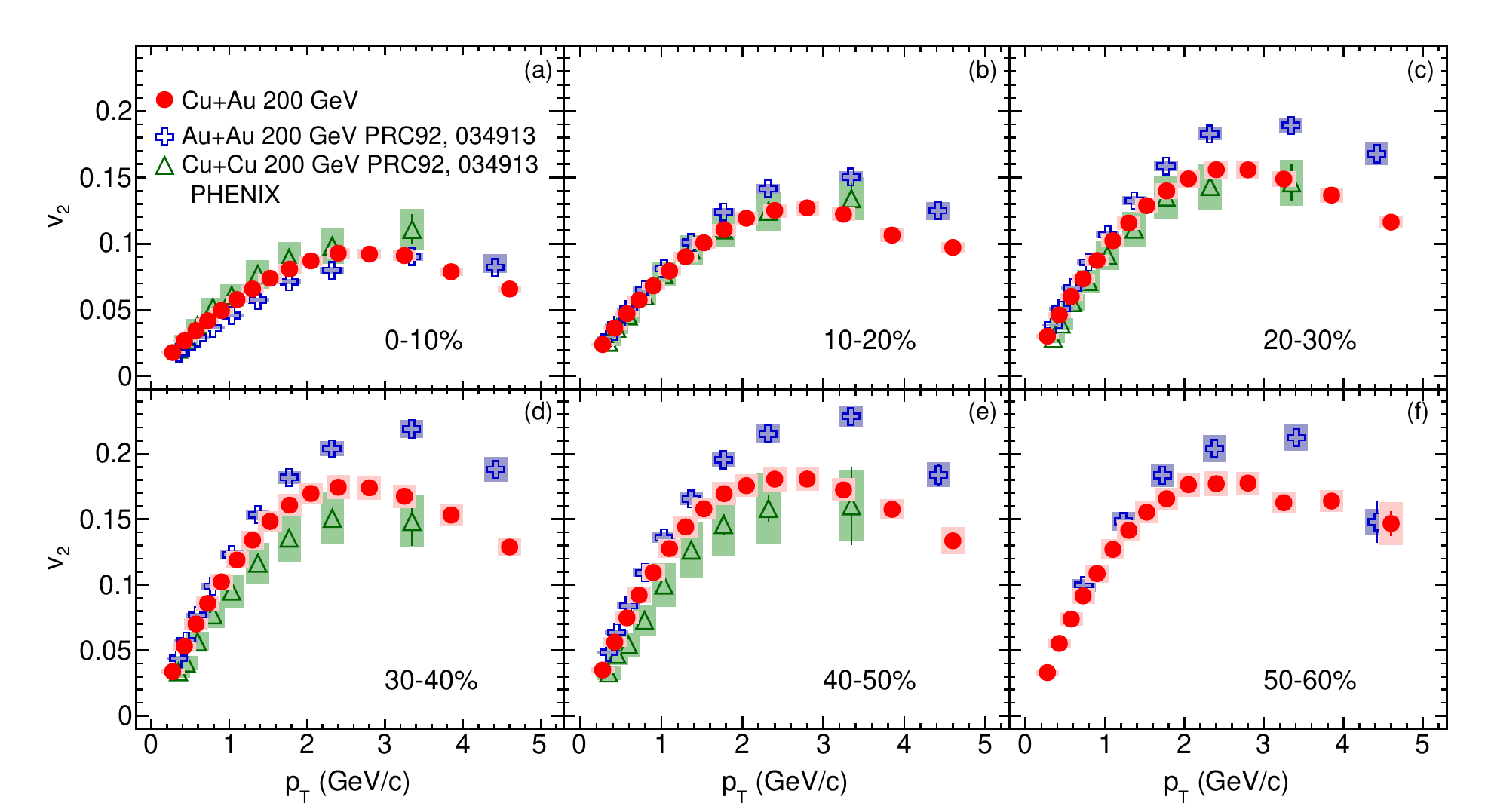}
\caption{
The second-order Fourier coefficients $v_2(p_T)$ for charged hadrons 
measured at midrapidity in Cu$+$Au, Au$+$Au~\cite{Adare:2014bga}, and 
Cu$+$Cu~\cite{Adare:2014bga} collisions at $\sqrt{s_{_{NN}}}$ = 200 GeV. In 
each panel, the $v_2(p_T)$ coefficients are compared for the same 
centrality class, as marked in the figure. The symbols represent the 
measured $v_2(p_T)$ values, the error bars show the statistical 
uncertainties, and the shaded boxes indicate the systematic uncertainties.
}
\label{fig:v2_sameCent}
\end{minipage}
\begin{minipage}{1.0\linewidth}
\includegraphics[width=1.0\linewidth]{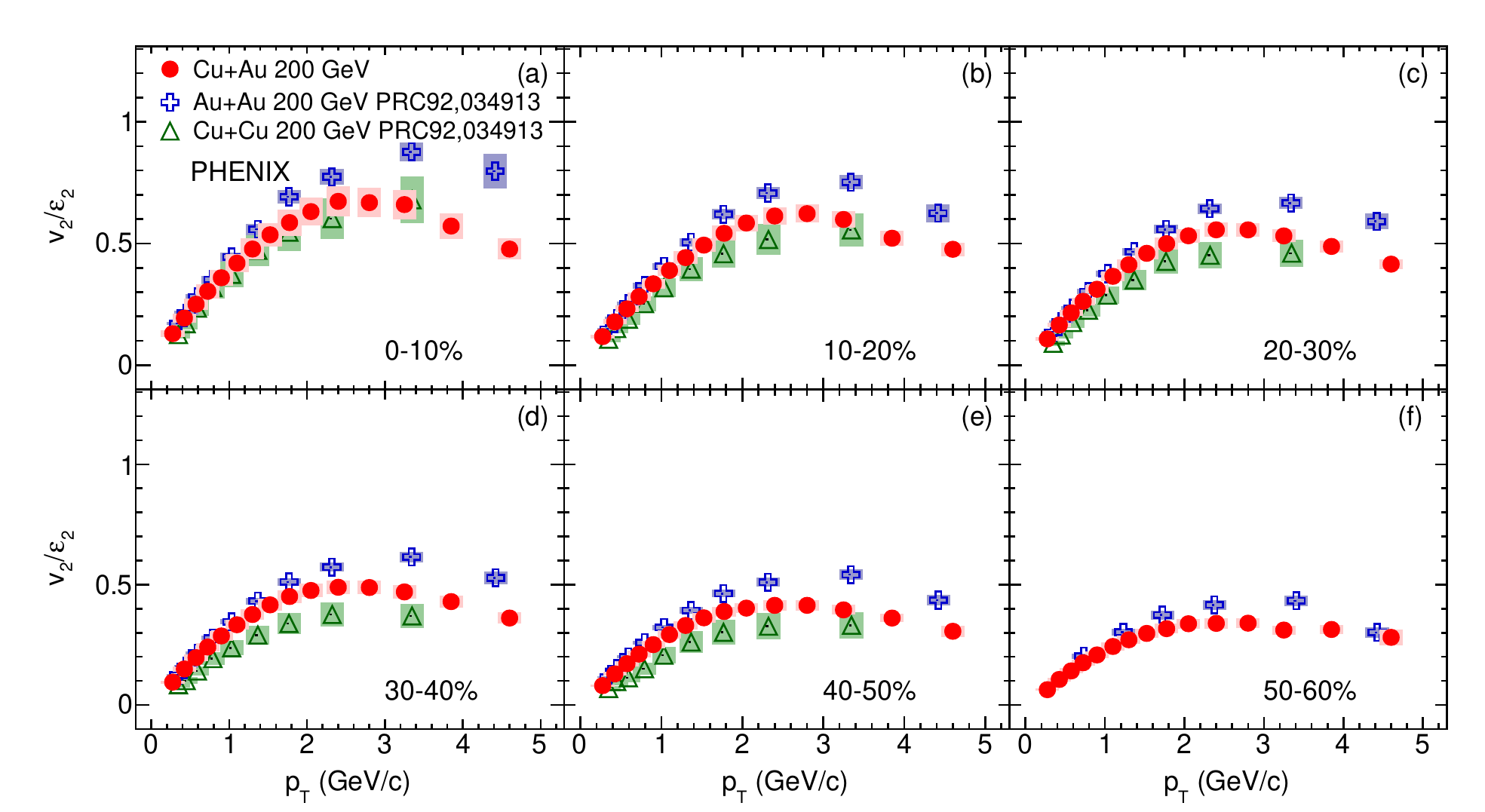}
\caption{
Scaled second-order Fourier coefficients $v_2(p_T)/\varepsilon_2$ for 
charged hadrons measured at midrapidity in 
Cu$+$Au,Au$+$Au~\cite{Adare:2014bga}, and Cu$+$Cu~\cite{Adare:2014bga} 
collisions at $\sqrt{s_{_{NN}}}$ = 200 GeV. In each panel, the $v_2(p_T)$ 
values measured in the centrality classes marked in the figure, are scaled 
by the average second-order participant eccentricity $\varepsilon_2$ in 
the initial state of the collisions as determined by a MC Glauber 
calculation described in the text. The symbols represent the scaled 
$v_2(p_T)/\varepsilon_2$ values, the error bars show the statistical 
uncertainties, and the shaded boxes indicate the systematic uncertainties.
}
\label{fig:v2Scaled_eps}
\end{minipage}
\end{figure*}

In Fig.~\ref{fig:v2_sameCent}, the $v_2(\pt)$ coefficients are compared 
for six different centrality selections. We observe that in each 
centrality class at a given \pt the values measured in Cu$+$Au collisions 
are always between those measured in Cu$+$Cu and Au$+$Au collisions. In 
all centrality classes chosen, the Cu$+$Cu system has larger elliptic 
eccentricity than both Cu$+$Au and Au$+$Au collisions. However, except in 
the most central 0\%--10\% collisions, the measured $v_2(\pt)$ values are 
not ordered according to the magnitude of $\varepsilon_2$ in the different 
systems listed in Table~\ref{tab:ecc_part_4systems}. To further 
investigate this, in Fig.~\ref{fig:v2Scaled_eps} we scale the $v_2(\pt)$ 
values in each collision system with their respective participant 
eccentricity $\varepsilon_2$. The resulting $v_2(\pt)/\varepsilon_2$ are 
ordered by system size, but this scaling does not lead to a universal 
behavior.

\begin{figure*}[htbp]
\includegraphics[width=1.0\linewidth]{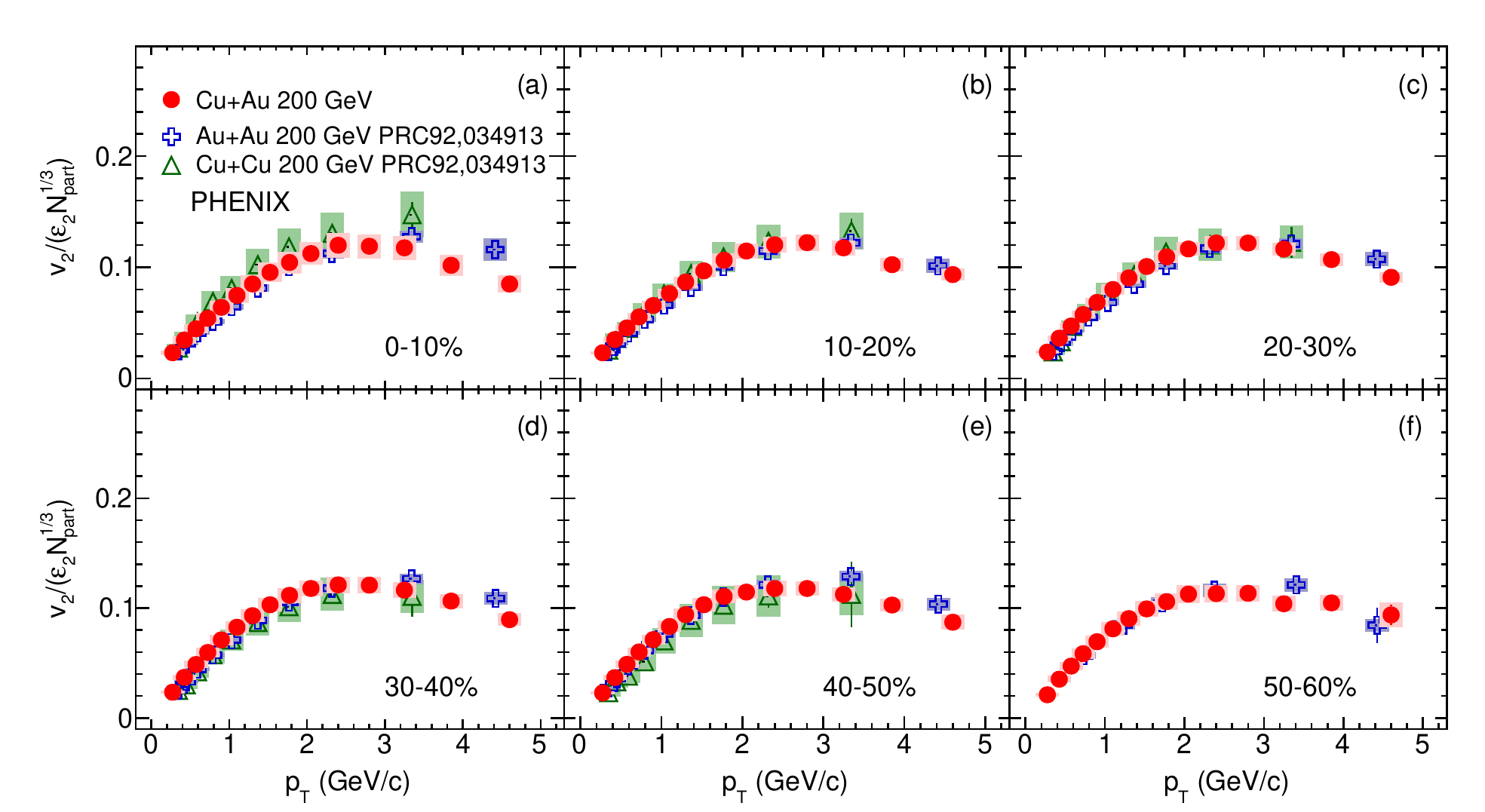}
\caption{
Scaled second-order Fourier coefficients $v_2(p_T)/(\varepsilon_2N_{\rm 
part}^{1/3})$ for charged hadrons measured at midrapidity in 
Cu$+$Au,Au$+$Au~\cite{Adare:2014bga}, and Cu$+$Cu~\cite{Adare:2014bga} 
collisions at $\sqrt{s_{_{NN}}}$ = 200 GeV. In each panel, the $v_2(p_T)$ 
values measured in the centrality classes marked in the figure, are scaled 
by the average second-order participant eccentricity $\varepsilon_2$ in 
the initial state of the collisions as determined by a MC Glauber 
calculation described in the text, and the corresponding number of nucleon 
participants $N_{\rm part}^{1/3}$. The symbols represent the scaled 
$v_2(p_T)/(\varepsilon_2N_{\rm part}^{1/3})$ values, the error bars show 
the statistical uncertainties, and the shaded boxes indicate the 
systematic uncertainties.
}
\label{fig:v2Scaled}
\end{figure*}

In Ref.~\cite{Adare:2014bga}, PHENIX compared measurements in Cu$+$Cu and 
Au$+$Au collisions for different center-of-mass energies and centrality 
selections and found that the $v_2$ values obey common empirical scaling 
with $\varepsilon_2N_{\rm part}^{1/3}$. The motivation for introducing the 
$N_{\rm part}^{1/3}$ factor is that under the assumption that $N_{\rm 
part}$ is proportional to the volume of the fireball, $N_{\rm part}^{1/3}$ 
is a quantity proportional to a length scale, and therefore may account 
for the system-size dependence of the $v_2$ values. In 
Fig.~\ref{fig:v2Scaled} we add to this comparison the results from the 
asymmetric Cu$+$Au collisions. This scaling brings the $v_2(\pt)$ results 
from the three collisions systems together across all centrality classes 
in this study.
   
\begin{figure*}[htbp]
\includegraphics[width=1.0\linewidth]{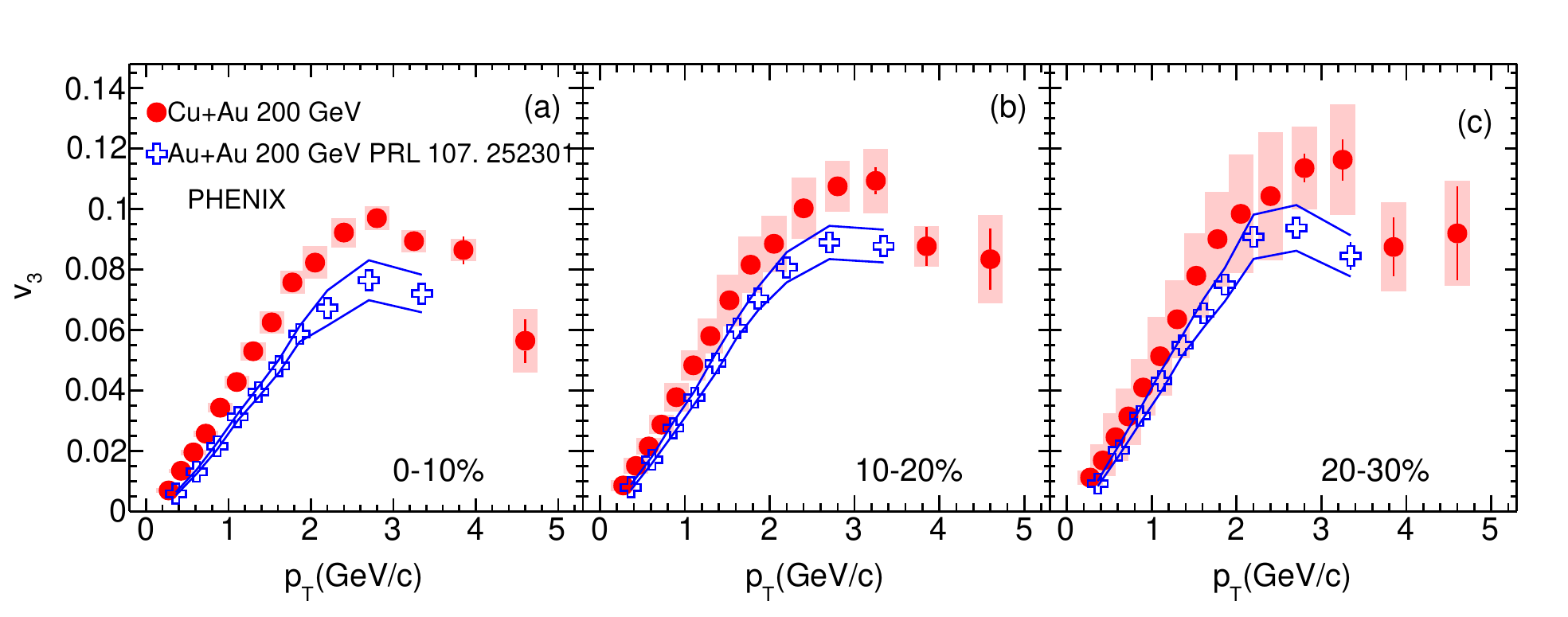}
\caption{
The third-order Fourier coefficients $v_3(p_T)$ for charged hadrons 
measured at midrapidity in Cu$+$Au and Au$+$Au~\cite{Adare:2011tg} 
collisions at $\sqrt{s_{_{NN}}}$ = 200 GeV. In each panel, the $v_3(p_T)$ 
coefficients are compared for the same centrality class, as marked in the 
figure. The symbols represent the measured $v_3(p_T)$ values, the error 
bars show the statistical uncertainties, and the shaded boxes indicate the 
systematic uncertainties.
}
\label{fig:v3_sameCent}
\end{figure*}

\begin{figure*}[htbp]
\begin{minipage}{1.0\linewidth}
\includegraphics[width=1.0\linewidth]{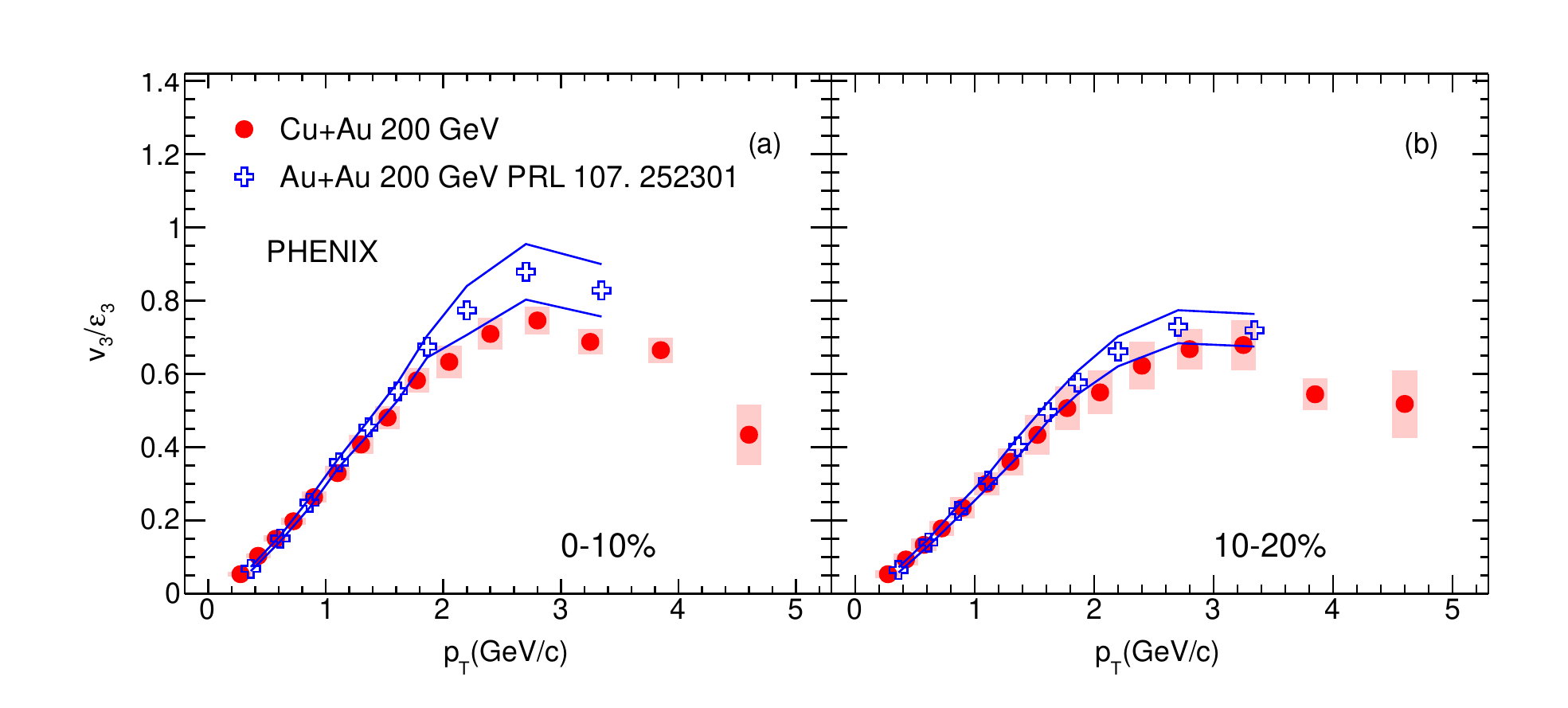}
\caption{
Scaled third-order Fourier coefficients $v_3(p_T)/\varepsilon_3$ for 
charged hadrons measured at midrapidity in Cu$+$Au and 
Au$+$Au~\cite{Adare:2011tg} collisions at $\sqrt{s_{_{NN}}}$ = 200 GeV. In 
each panel, the $v_3(p_T)$ values measured in the centrality classes 
marked in the figure, are scaled by the average third-order participant 
eccentricity $\varepsilon_3$ in the initial state of the collisions as 
determined by a MC Glauber calculation described in the text. The symbols 
represent the scaled $v_3(p_T)/\varepsilon_3$ values, and the error bars 
show the statistical uncertainties. The shaded boxes indicate the 
systematic uncertainties in the Cu$+$Au measurements, and the lines around 
the points marked with a cross show the systematic uncertainties in the 
Au$+$Au measurements.
}
\label{fig:v3Scaled_eps3}
\end{minipage}
\begin{minipage}{1.0\linewidth}
\includegraphics[width=1.0\linewidth]{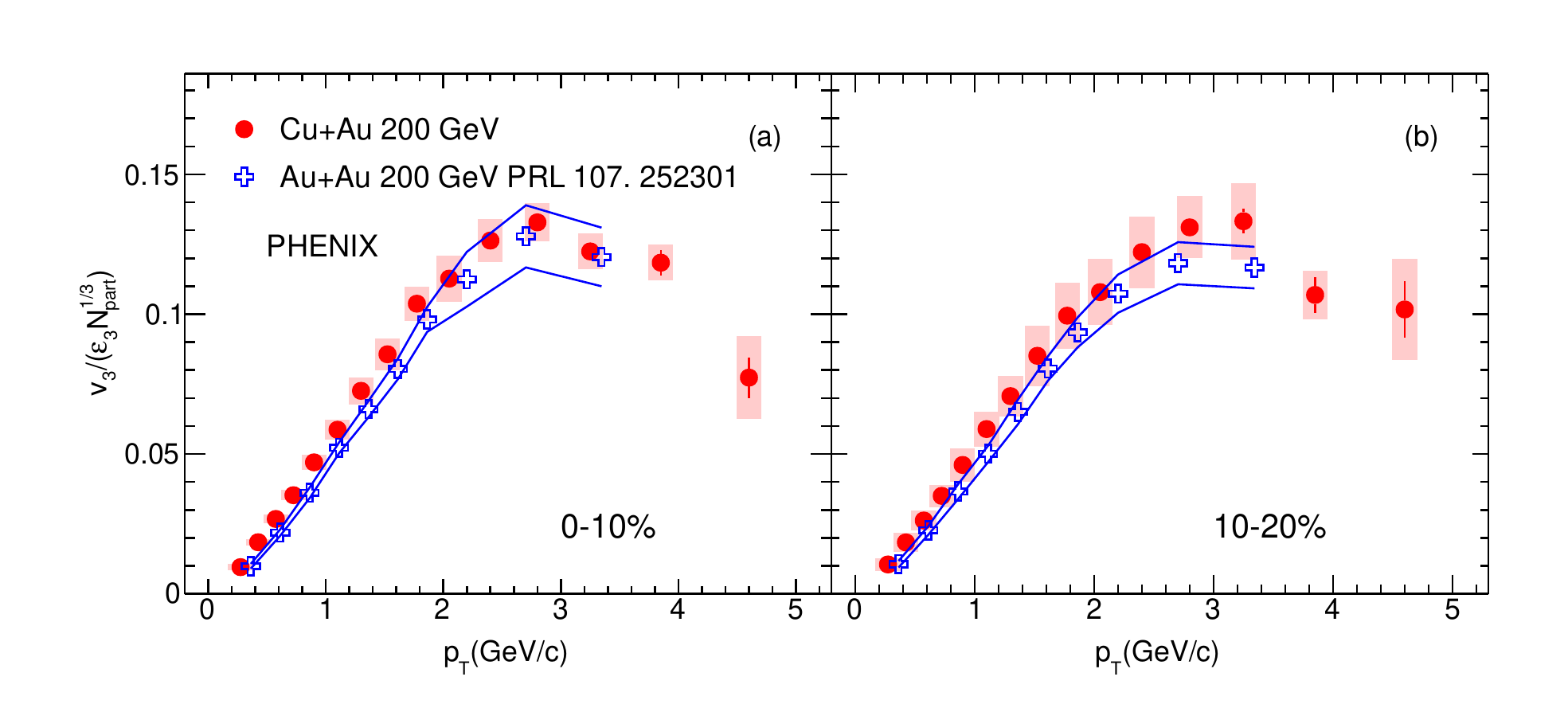}
\caption{
Scaled third-order Fourier coefficients $v_3(p_T)/(\varepsilon_3N_{\rm 
part}^{1/3})$ for charged hadrons measured at midrapidity in Cu$+$Au and 
Au$+$Au~\cite{Adare:2011tg} collisions at $\sqrt{s_{_{NN}}}$ = 200 GeV. In 
each panel, the $v_3(p_T)$ values measured in the centrality classes 
marked in the figure, are scaled by the average third-order participant 
eccentricity $\varepsilon_3$ in the initial state of the collisions as 
determined by a MC Glauber calculation described in the text, and the 
corresponding number of nucleon participants $N_{\rm part}^{1/3}$. The 
symbols represent the scaled $v_3(p_T)/(\varepsilon_3N_{\rm part}^{1/3})$ 
values, and the error bars show the statistical uncertainties. The shaded 
boxes indicate the systematic uncertainties in the Cu$+$Au measurements, 
and the lines around the points marked with a cross show the systematic 
uncertainties in the Au$+$Au measurements.
}
\label{fig:v3Scaled}
\end{minipage}
\end{figure*}

In Fig.~\ref{fig:v3_sameCent} the $v_3(\pt)$ values are compared in 
Cu$+$Au and Au$+$Au collisions for events of the same centrality. Unlike 
in the $v_2(\pt)$ measurements, here the values of $v_3(\pt)$ are ordered 
according to the initial triangularities $\varepsilon_3$ listed in 
Table~\ref{tab:ecc_part_4systems}, with the Cu$+$Au results being larger 
than the Au$+$Au ones. In particular, in the most central 0\%--10\% 
collisions $\varepsilon_3$ in Cu$+$Au is about 50\% larger than in Au$+$Au 
collisions, and a similar difference is observed in the $v_3(\pt)$ values. 
In Fig.~\ref{fig:v3Scaled_eps3} the $v_3(\pt)$ values are scaled by the 
initial $\varepsilon_3$ eccentricity. A good agreement between the 
different systems is observed at low \pt ($\le 2$ GeV/$c$), which 
indicates that the participant eccentricities obtained in the Glauber 
model provide an adequate description of the fluctuating initial geometry. 
Additionally, we perform scaling with $\varepsilon_3N_{\rm part}^{1/3}$, 
as was done for the $v_2(\pt)$ measurements. The results of this scaling 
are shown in Fig.~\ref{fig:v3Scaled}.  In this case, the measurement in 
Cu$+$Au and Au$+$Au collisions are in better agreement at high \pt, 
however at low \pt the $v_3(\pt)/\varepsilon_3N_{\rm part}^{1/3}$ values 
are systematically higher for the Cu$+$Au system.

\vspace{-0.1cm}


	\subsection{Theory comparisons}
	\label{sec:theory}

   \subsubsection{Hydrodynamic calculations}
   \label{sec:theoryHydro}

Predictions from 3D+1 viscous hydrodynamic calculations are 
available~\cite{Bozek:2012hy}. At low \pt ($ < 1.0 $ GeV/$c$) directed 
flow is predicted to be in the hemisphere of the Cu side, while for high 
\pt ( $ > 1.5 $ GeV/$c$) directed flow is predicted to be in the 
hemisphere on the Au side. Further, the bulk directed flow component from 
integration over \pt is predicted to be in the Cu-nucleus hemisphere.  
Due to the large systematic uncertainties and small value of $v_1$ at 
small \pt, we can not reliably determine the sign of the $v_1$ component 
at low \pt, or the sign of the bulk directed flow. At high \pt the 
measurement is in agreement with the directed flow being in the Au 
hemisphere, under the assumption that the spectator neutrons are deflected 
outward from the interaction region and aligned with the impact parameter 
vector. Ref.~\cite{Bozek:2012hy} shows the $v_1$ with respect to the 
reaction plane (i.e. the impact parameter vector) for 20\%--30\% central 
Cu$+$Au collisions including particles within $|\eta|<1.0$, and thus we 
cannot compare directly with our narrower rapidity selection.  It is 
notable however, that the hydrodynamic results at \pt=~2~GeV/$c$ reach 
$v_{1} \approx $ 5\%, while the experimental data within $|\eta|<0.35$ are 
less than 2\%. 

\begin{figure*}[htbp]
\begin{minipage}{1.0\linewidth}
\includegraphics[width=1.0\linewidth]{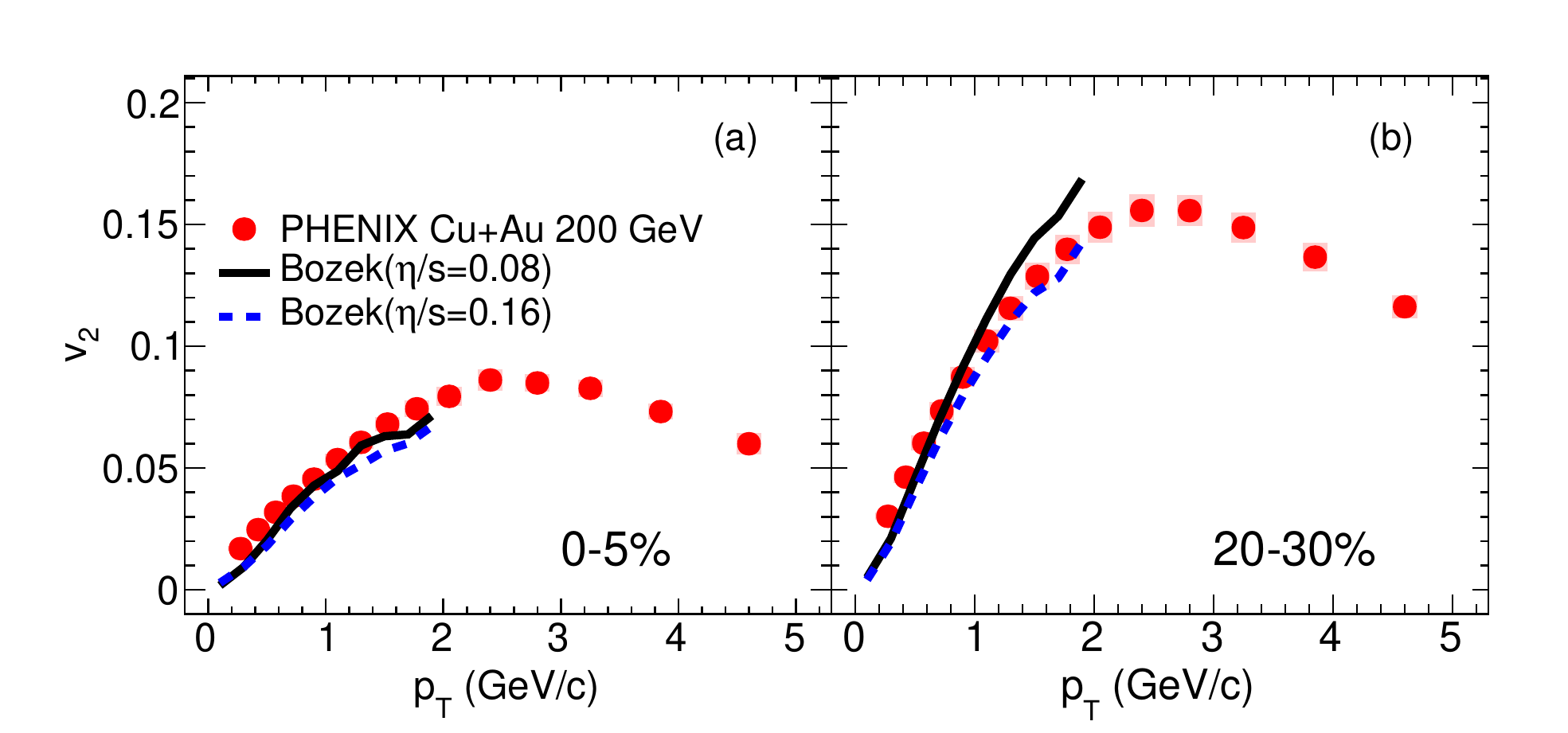}
\caption{
The second-order Fourier coefficients $v_2(p_T)$ for charged hadrons 
measured at midrapidity in Cu$+$Au collisions at $\sqrt{s_{_{NN}}}$ = 200 GeV 
in comparison to hydrodynamics calculations for the centrality classes 
marked in each panel. The symbols represent the measured $v_2(p_T)$ 
values, the error bars show the statistical uncertainties, and the shaded 
boxes indicate the systematic uncertainties. The theoretical calculations, 
shown with the solid and dashed lines, are performed with two different 
values of the specific viscosity $\eta/s$ marked in the figure.
}
\label{fig:v2_hydro}
\end{minipage}
\begin{minipage}{1.0\linewidth}
\includegraphics[width=1.0\linewidth]{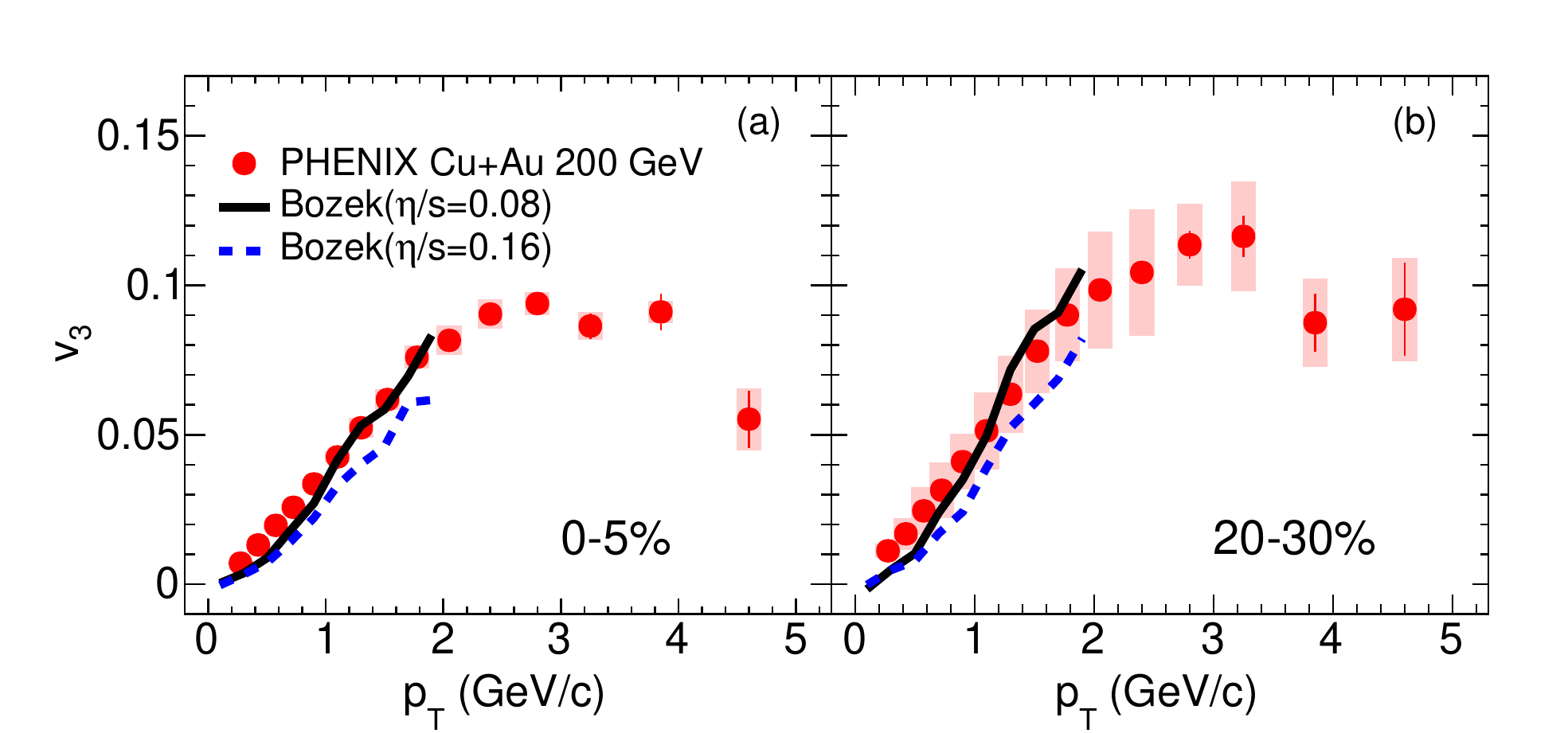}
\caption{
The third-order Fourier coefficients $v_3(p_T)$ for charged hadrons 
measured at midrapidity in Cu$+$Au collisions at 
$\sqrt{s_{_{NN}}}=200$~GeV in comparison to hydrodynamics calculations for 
the centrality classes marked in each panel. The symbols represent the 
measured $v_3(p_T)$ values, the error bars show the statistical 
uncertainties, and the shaded boxes indicate the systematic uncertainties. 
The theoretical calculations, shown with the solid and dashed lines, are 
performed with two different values of the specific viscosity $\eta/s$ 
marked in the figure.
}
\label{fig:v3_hydro}
\end{minipage}
\end{figure*}

The predictions for elliptic and triangular flow as a 
function of \pt are compared to the data in Fig.~\ref{fig:v2_hydro} and 
Fig.~\ref{fig:v3_hydro}. Calculations with two different values of the 
specific viscosity ${\eta/s}=0.08$ and ${\eta/s}=0.16$ are shown. Our 
measurements in the 20\%--30\% centrality range are consistent with 
each of these values; for the most central 0\%--5\% events, a value of 
${\eta/s}=0.08$ is closer to the data.

	\subsubsection{AMPT}
	\label{sec:theoryAMPT}

The A-Multiphase-Transport Model (AMPT) 
generator~\cite{Lin:2004en,Chen:2005zy} has been established as a useful 
tool in the study of flow observables in heavy-ion collisions 
~\cite{Koop:2015wea}. Therefore, it is of interest to compare the measured 
$v_1,v_2,$ and $v_3$ as a function of $p_T$ with the corresponding 
quantities calculated using the AMPT model. To that end, we used AMPT 
v2.21 with string melting turned on to generate approximately 2 million 
minimum bias Cu$+$Au events at $\sqrt{s_{_{NN}}}=200$ GeV, setting the partonic 
cross section alternately to $\sigma_{\rm part} = 1.5$ mb and 3.0 mb. In the 
default version of the model, initial conditions are generated using Monte 
Carlo Glauber with a \emph{gray disk} approach to nucleon-nucleon 
interactions. However, in this study we utilize a modified \emph{black 
disk} Glauber model with a fixed nucleon-nucleon inelastic cross section 
of 42 mb, as used in Ref.~\cite{Koop:2015wea}.

Following the method of \cite{Koop:2015wea}, Fourier coefficients $v_1, 
v_2,$ and $v_3$ are calculated for unidentified charged hadrons within 
$|\eta| < 0.35$, with respect to the corresponding participant planes 
$\Psi_1, \Psi_2,$ and $\Psi_3$. These plane angles are computed for each 
event from the initial coordinates of nucleon participants with a Gaussian 
smearing of width $\sigma=0.4$ fm.

\begin{figure*}[htbp]
\begin{minipage}{1.0\linewidth}
\includegraphics[width=1.0\linewidth]{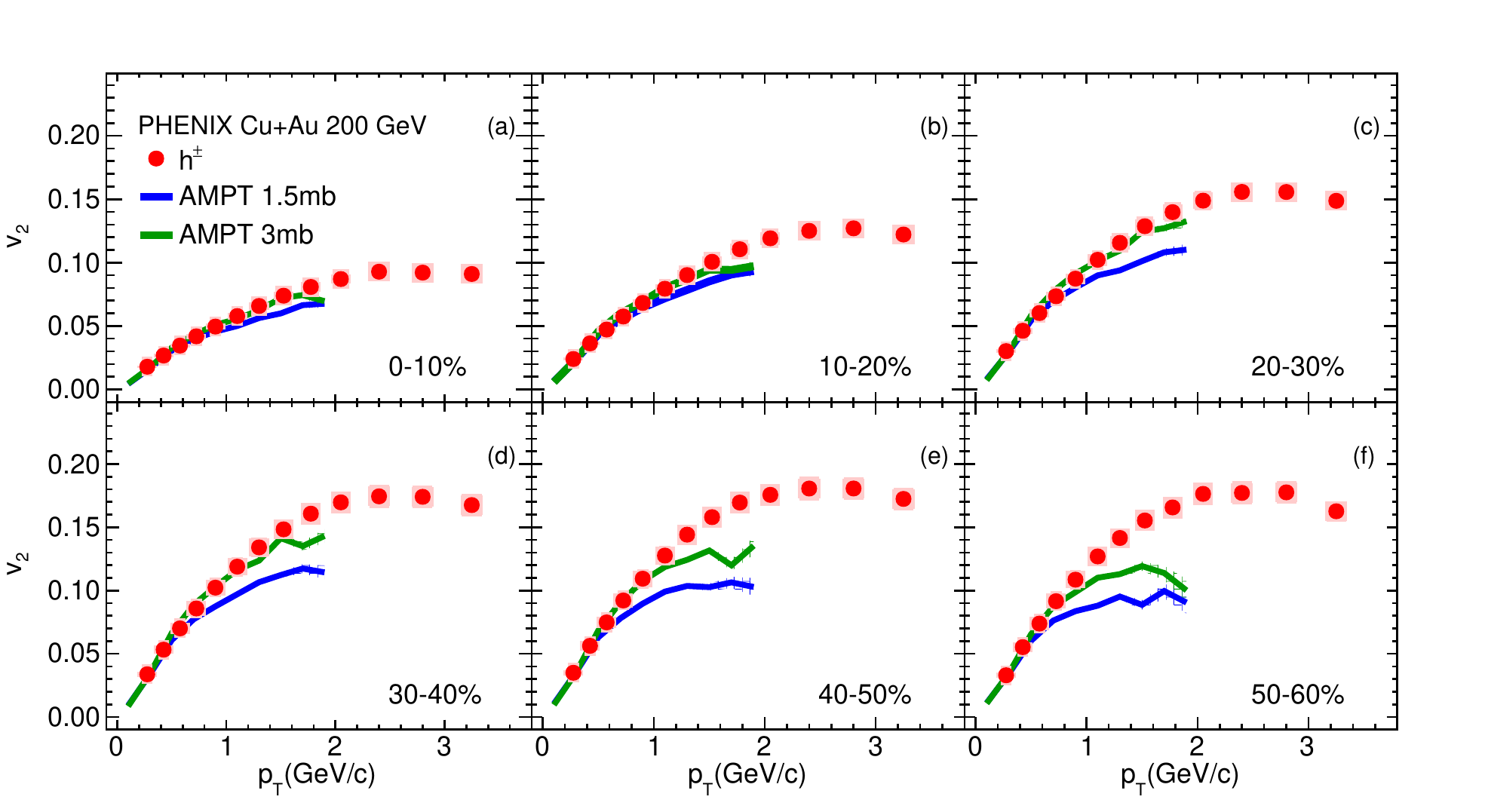}
\caption{
The second-order Fourier coefficients $v_2(p_T)$ for charged hadrons 
measured at midrapidity in Cu$+$Au collisions at $\sqrt{s_{_{NN}}}$ = 200 GeV 
in comparison to AMPT model calculation for the centrality classes marked 
in each panel. The symbols represent the measured $v_2(p_T)$ values, the 
error bars show the statistical uncertainties, and the shaded boxes 
indicate the systematic uncertainties. For the theoretical calculations, 
which are shown with lines, only statistical uncertainties are plotted.
}
\label{fig:v2_AMPT}
\end{minipage}
\begin{minipage}{1.0\linewidth}
\includegraphics[width=1.0\linewidth]{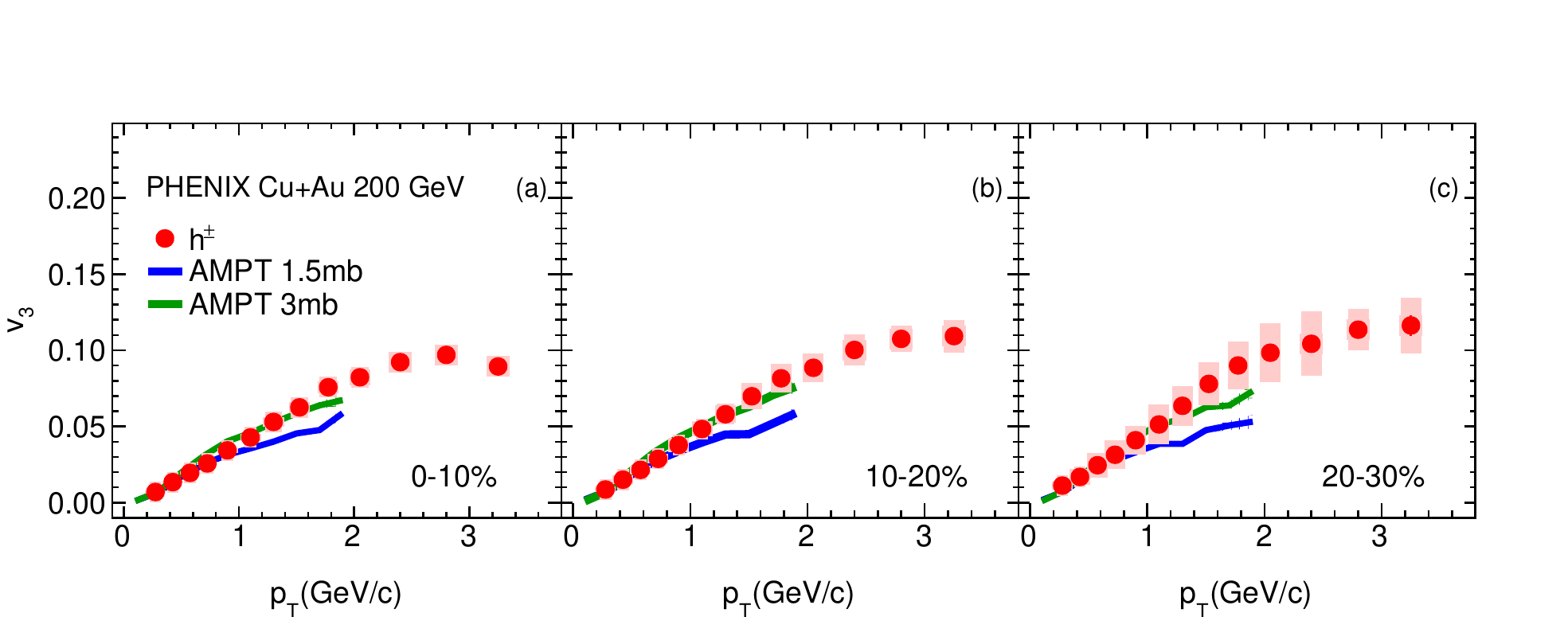}
\caption{
The third-order Fourier coefficients $v_3(p_T)$ for charged hadrons 
measured at midrapidity in Cu$+$Au collisions at $\sqrt{s_{_{NN}}}$ = 200 GeV 
in comparison to AMPT model calculation for the centrality classes marked 
in each panel. The symbols represent the measured $v_3(p_T)$ values, the 
error bars show the statistical uncertainties, and the shaded boxes 
indicate the systematic uncertainties. For the theoretical calculations, 
which are shown with lines, only statistical uncertainties are plotted.
}
\label{fig:v3_AMPT}
\end{minipage}
\end{figure*}

The $v_2(p_T)$ and $v_3(p_T)$ results shown in Fig.~\ref{fig:v2_AMPT} and 
Fig.~\ref{fig:v3_AMPT} are well reproduced by the model for $p_T<1 $ 
GeV/$c$. The comparison with the data indicates that the 3.0 mb partonic 
cross section gives a better description of the system dynamics.
However, the calculation of $v_1$ and its comparison with experimental 
data is less straightforward.  Because the experimentally measured Cu 
spectator neutron orientation is unknown, we calculate the $v_1$ values 
with respect to the impact parameter vector $\vec{b}$ pointing in the 
direction of the Cu nucleus as well as with respect to $\Psi_1$, the 
overlap region calculated as previously described. Because the calculation 
is done in the participant center-of-mass frame, weighting all 
participants equally yields exactly $\varepsilon_1$ = 0 and hence no 
direction for $\Psi_1$. There are various suggestions in the literature 
for weighting with $r^2$ and $r^3$ ~\cite{Gardim:2011qn,Teaney:2010vd}, 
and in this study we choose to use $r^2$.

\begin{figure*}[htbp]
\includegraphics[width=1.0\linewidth]{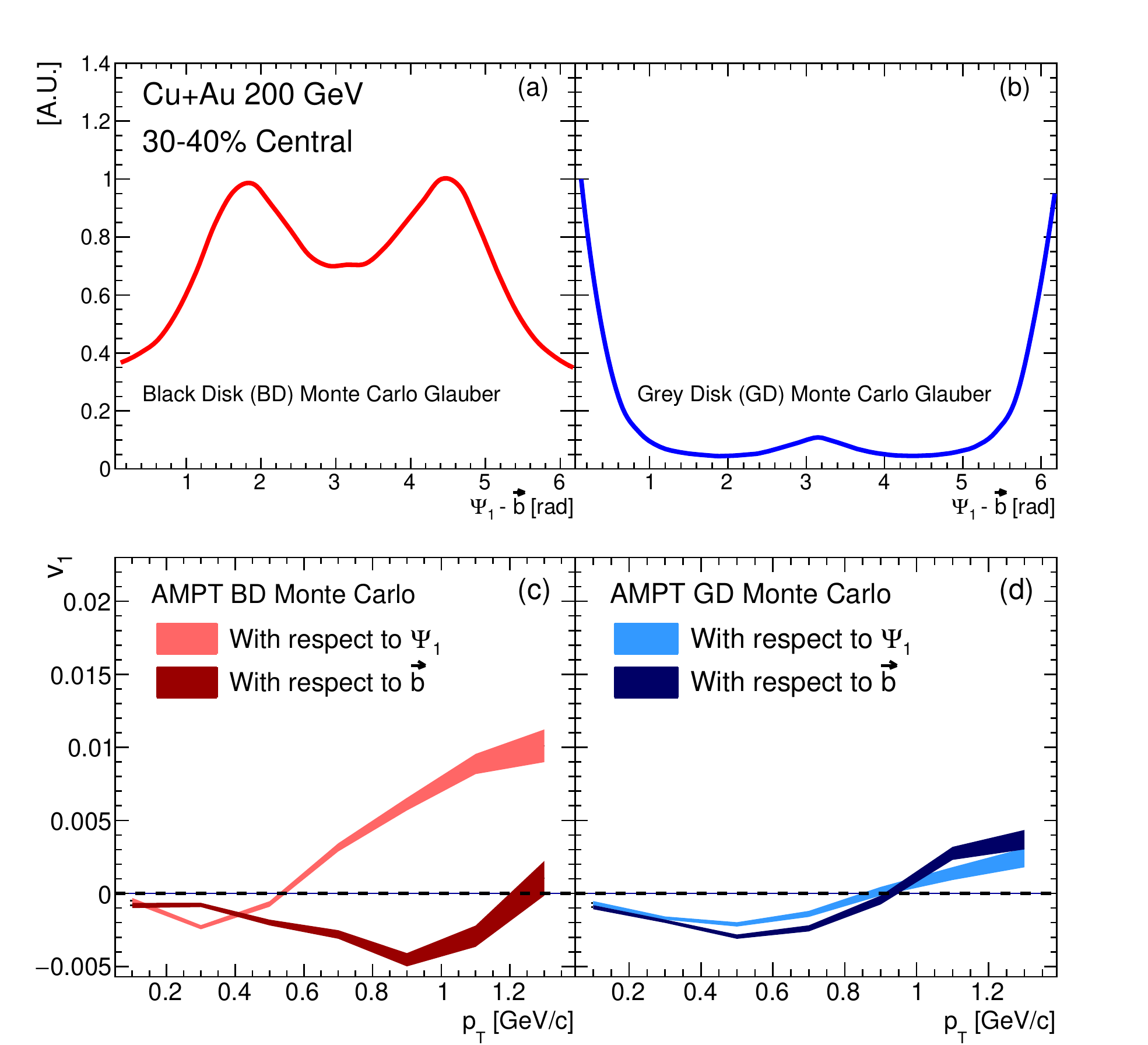}
\caption{
The top panels show the maximum-normalized distribution of first-order 
participant plane angle $\Psi_1$ computed from the initial coordinates of 
participant nucleons determined with (a) black disk, and (b) gray disk 
Monte Carlo Glauber simulations. The bottom panels show AMPT $v_1$ 
computed with respect to the impact parameter $\vec{b}$, and $\Psi_1$ 
using (c) black disk and (d) gray disk Monte Carlo for the initial 
conditions. }
\label{fig:AMPT_Glauber_Comparison}
\end{figure*}

In addition, we have considered two different Monte Carlo Glauber initial 
conditions, one with black disk (BD) nucleons and one with gray disk (GD) 
nucleons, thus varying the diffuseness of the nucleon-nucleon interaction 
radius. Figure~\ref{fig:AMPT_Glauber_Comparison} shows results for Cu$+$Au 
collisions within the 30\%-40\% centrality selection on the relative 
distribution of $\Psi_1$ to $\vec{b}$ pointing the direction of the Cu 
nucleus.  Panel (a) is for the BD case and Panel (b) the GD case.  This 
small difference in the treatment of initial geometry completely 
re-orients the $\Psi_1$ vector.  The lower panels show the AMPT 
midrapidity particle $v_1$ as a function of \pt relative to $\Psi_1$ and 
$\vec{b}$ again the BD and GD implementation.  It is interesting to note 
that in the GD case where the two results agree, the prediction is for low 
\pt particles moving in the direction of the Au nucleus and the high \pt 
particles in the direction of the Cu nucleus (opposite to the previously 
discussed hydrodynamic prediction).

\clearpage

We note that it is currently unknown whether the spectator 
neutrons bend toward or away from the interaction overlap region between 
the nuclei, and whether they are oriented along the impact parameter 
vector $\vec{b}$, along the vector $\Psi_1$ determined by the initial 
energy density in the overlap region, or some other vector.  In fact, it 
is conceivable that spectators very close to the overlap region have a 
different behavior from spectators far away from the overlap.  These 
ambiguities need resolution before a more direct theory to data comparison 
can be made.


 		\section{Summary}
		\label{sec:summary}

Anisotropic flow coefficients for inclusive charged particles and 
identified hadrons $\pi^{\pm}$, $K^{\pm}$, $p$, and $\bar{p}$ produced in 
Cu$+$Au collisions at $\sqrt{s_{_{NN}}}$ = 200 GeV have been measured by the 
PHENIX experiment at RHIC using event plane techniques. The $v_1$, $v_2$, 
and $v_3$ measurements were performed at midrapidity as a function of 
transverse momentum \pt over a broad range of collision centralities. Mass 
ordering was observed for low \pt in the identified particle measurements, 
as predicted by hydrodynamics.

A system size comparison was performed for the inclusive charged particles 
using previous PHENIX measurements at $\sqrt{s_{_{NN}}}$ = 200 GeV of 
$v_2(p_T)$ in Cu$+$Cu and Au$+$Au collisions, and $v_3(p_T)$ in Au$+$Au 
collisions. The elliptic and triangular flow measurements between different systems and 
centrality selections were found to scale with the product of the initial 
participant eccentricity and the third root of the number of nucleon 
participants $\varepsilon_nN_{\rm part}^{1/3}$. The system size dependence 
of the $v_3(p_T)$ values could also be  described by participant eccentricity 
$\varepsilon_3$ scaling alone.

The inclusive charged-particle measurements were compared to theoretical 
predictions. In the $v_1$ measurement, we observed negative values at high 
\pt, indicating that hadrons are emitted in the transverse plane 
preferentially in the hemisphere of the spectators from the Au nucleus, 
assuming that they moved outward from the interaction region and are 
aligned with the impact parameter vector. The AMPT transport model 
calculations were found to be in agreement with the magnitude of the 
measured $v_1(p_T)$ signals, but having the opposite sign. At low \pt ($< 
1 $ GeV/$c$) AMPT provides a reasonable description of the triangular flow 
in all measured centrality classes that cover the 0\%--30\% range, and the 
elliptic flow measurements in the 0\%--60\% range.  Event-by-event 
hydrodynamics calculations with specific viscosity in the range $\eta/s = 
0.08-0.16$ reproduce the measured $v_2(p_T)$ and $v_3(p_T)$ values.


\section*{ACKNOWLEDGMENTS}   

We thank the staff of the Collider-Accelerator and Physics
Departments at Brookhaven National Laboratory and the staff of
the other PHENIX participating institutions for their vital
contributions.  We acknowledge support from the
Office of Nuclear Physics in the
Office of Science of the Department of Energy,
the National Science Foundation,
Abilene Christian University Research Council,
Research Foundation of SUNY, and
Dean of the College of Arts and Sciences, Vanderbilt University
(U.S.A),
Ministry of Education, Culture, Sports, Science, and Technology
and the Japan Society for the Promotion of Science (Japan),
Conselho Nacional de Desenvolvimento Cient\'{\i}fico e
Tecnol{\'o}gico and Funda\c c{\~a}o de Amparo {\`a} Pesquisa do
Estado de S{\~a}o Paulo (Brazil),
Natural Science Foundation of China (P.~R.~China),
Croatian Science Foundation and
Ministry of Science, Education, and Sports (Croatia),
Ministry of Education, Youth and Sports (Czech Republic),
Centre National de la Recherche Scientifique, Commissariat
{\`a} l'{\'E}nergie Atomique, and Institut National de Physique
Nucl{\'e}aire et de Physique des Particules (France),
Bundesministerium f\"ur Bildung und Forschung, Deutscher
Akademischer Austausch Dienst, and Alexander von Humboldt Stiftung 
(Germany),
National Science Fund, OTKA, K\'aroly R\'obert University College,
and the Ch. Simonyi Fund (Hungary),
Department of Atomic Energy and Department of Science and Technology 
(India),
Israel Science Foundation (Israel),
Basic Science Research Program through NRF of the Ministry of Education 
(Korea),
Physics Department, Lahore University of Management Sciences (Pakistan),
Ministry of Education and Science, Russian Academy of Sciences,
Federal Agency of Atomic Energy (Russia),
VR and Wallenberg Foundation (Sweden),
the U.S. Civilian Research and Development Foundation for the
Independent States of the Former Soviet Union,
the Hungarian American Enterprise Scholarship Fund,
and the US-Israel Binational Science Foundation.




%
 
\end{document}